\documentclass[11pt,twoside]{article}
\usepackage{mathtext,amssymb,amsmath}
\usepackage[cp1251]{inputenc}
\usepackage[T2A]{fontenc}
\usepackage[english,russian]{babel}
\usepackage{graphicx}
\usepackage{cite}
\usepackage{authblk}
\usepackage{color}

\begin{document}
	
\newcommand {\be}{\begin{equation}}
\newcommand {\ee}[1] {\label{#1} \end{equation}}
\newcommand {\e} {\varepsilon}
\newcommand {\ph} {\varphi}
\newcommand {\la} {\langle}
\newcommand {\ra} {\rangle}
\newcommand {\lla} {\left\langle}
\newcommand {\rra} {\right\rangle}
\renewcommand {\d} {\mathrm{d}}
\renewcommand {\Re} {\mathrm{Re}}
\renewcommand {\Im} {\mathrm{Im}}
\renewcommand{\thefootnote}{\arabic{footnote}}

УДК 537.86, 001.891.57, 621.37

\begin{center}
    {\bf СИНХРОНИЗАЦИЯ В АНСАМБЛЯХ КУРАМОТО-САКАГУЧИ ПРИ КОНКУРИРУЮЩЕМ ВЛИЯНИЯ ОБЩЕГО ШУМА И ГЛОБАЛЬНОЙ СВЯЗИ}

\vspace{5mm}
    {\it Д.\ С.\ Голдобин$^{1,2}$, А.\ В.\ Долматова$^1$, М.\ Розенблюм$^{3,4}$, А.\ Пиковский$^{3,4}$}

\vspace{5mm}
$^1$Институт механики сплошных сред УрО РАН\\
614013 Пермь, ул.~Акад.~Королева, 1\\
$^2$Пермский государственный национальный исследовательский университет\\
614990 Пермь, ул.~Букирева, 15\\
$^3$University of Potsdam\\
14476 Potsdam-Golm, Karl-Liebknecht-Str., 24/25\\
$^4$Нижегородский государственный университет им.\ Н.~И.\ Лобачевского\\
603950 Нижний Новгород, пр.~Гагарина, 23
\end{center}


\begin{abstract}
В работе исследуются эффекты синхронизации и десинхронизации в ансамблях фазовых осцилляторов с глобальной связью типа Курамото-Сакагучи при воздействии на них общим шумом. В связи с тем, что механизмы синхронизации за счет связи и общего шума существенно различны, представляет интерес выяснение особенностей их взаимодействия. В термодинамическом пределе большого числа осцилляторов, с помощью подхода Отта-Антонсена, выведены стохастические уравнения для параметра порядка и изучена их динамика как в случае идентичных осцилляторов, так и в случае малой расстройки собственных частот. Для идентичных осцилляторов исследована устойчивость состояния полной синхронизации и выявлено, что достаточный уровень общего шума может синхронизировать систему даже при отрицательной (отталкивающей) глобальной связи. Установлено нарушение равноправия между состояниями максимальной асинхронности (нулевого значения параметра порядка) и состоянием полной синхронизации: первое может быть только слабо притягивающим, тогда как второе может становиться адсорбирующим (переход к синхронизации становится необратимым). Исследована динамика перехода в синхронное состояние в зависимости от параметров. Для неидентичных осцилляторов полная синхронизация невозможно и адсорбирующее состояние исчезает: на его месте остается слабо притягивающее. Обнаружен и исследован нетривиальный эффект расхождения индивидуальных частот осцилляторов с отличающимися собственными частотами при умеренной отталкивающей связи, причем параметр порядка в этом случае остается достаточно большим. В Приложении к работе дается введение в теории Отта-Антонсена и Ватанабе-Строгаца.

\noindent{\it Ключевые слова:} Синхронизация, стохастические процессы, ансамбль Курамото-Сакагучи, подход Отта-Антонсена.

\end{abstract}

\section*{Введение}
Явление синхронизации в ансамблях осцилляторов с разными видами связей хорошо изучено и освещено в литературе~\cite{Pikovsky-Rosenblum-Kurths-2001,Crawford-1994}. Эффект синхронизации оказывает существенное влияние на поведение различных физических систем, например, лазерных установок или сверхпроводящих джозефсовновских контактов.
 Яркой иллюстрацией действия синхронизации в инженерных системах может служить инцидент с мостом Миллениум в Лондоне, который был закрыт через два дня после открытия, так как при прохождении большого количества людей конструкция начинала ощутимо раскачиваться из стороны в сторону~\cite{Strogatz-etal-2005}.
Не менее важную роль описываемый эффект играет и в биологических системах, в частности, при некоторых нейродегенеративных заболеваниях наблюдается патологическая синхронизация активности нейронов~\cite{Golomb-Hansel-Mato-2001}. Более того, синхронизацию можно наблюдать во многих социальных системах.

Еще одним возможным механизмом синхронизации осцилляторов является воздействие на них общим шумом~\cite{Pikovsky-1984}. Такой вид синхронизации также наблюдается в самых разных системах, в том числе оптических~\cite{Uchida-McAllister-Roy-2004}, экологических~\cite{Grenfell-etal-1998} и нейронных~\cite{Mainen-Sejnowski-1995}. Несмотря на то, что в последние годы математическая теория синхронизации общим шумом получила значительное развитие~\cite{Ritt-2003,Goldobin-Pikovsky-2004,Teramae-Tanaka-2005,Goldobin-Pikovsky-2005a,	Goldobin-Pikovsky-2005b,Goldobin-Pikovsky-2006,Malyaev-etal-2007,Wieczorek-2009, Goldobin-etal-2010,Goldobin-2014,Goldobin-2014b}, физический механизм этого эффекта все еще остается не столь очевидным, как механизм синхронизации при наличии связи (см., например,~\cite{Goldobin-Pikovsky-2005b,Goldobin-Pikovsky-2006,Braun-etal-2012}).

Примечательно, что воздействие малым общим шумом на ансамбль осцилляторов может иметь только синхронизирующее воздействие, тогда как общая связь может как синхронизировать, так и десинхронизировать систему (другими словами, она может быть как притягивающей, так и отталкивающей). Изучение взаимодействия влияний связи и общего шума на свойства синхронизации системы особенно интересно в связи с тем, что физические механизмы, лежащие в основе этих явлений, совершенно различны. Десинхронизация отталкивающей связью противодействует синхронизации общим шумом нетривиальным образом: они не могут компенсировать друг друга. Понимание тонких механизмов этого взаимодействия практически востребовано в тех случаях, когда необходимо противодействовать синхронизации, к которой приводит наличие общего шума, путем введения взаимной связи, что актуально для многих технических и биологических систем.

Некоторое время назад, в работе~\cite{Pimenova-etal-2016}, были описаны нетривиальные эффекты, возникающие при воздействии на ансамбль идентичных осцилляторов Курамото синхронизирующим общим шумом в присутствии десинхронизирующей глобальной связи; полученные результаты  дополняют более ранние работы~\cite{Garcia-Alvarez-etal-2009,Nagai-Kori-2010}. Было показано, что при наличии умеренной отталкивающей связи синхронизирующий эффект общего шума оказывается преобладающим, и система приходит в синхронное состояние. Однако при наличии расстройки собственных частот осцилляторов, средние частоты отдельных осцилляторов не притягиваются, а, наоборот, расталкиваются. Эффект расталкивания частот при синхронизации особенно примечателен в контексте того, что, например, Н.~Винер (N.~Wiener) определял синхронизацию, как ``phenomenon of the pulling together of frequencies'' (``явление взаимного притяжения частот'')~\cite{Wiener-1965}. При наличии синхронизирующей связи полный захват частоты не происходит, хотя средние частоты отдельных осцилляторов притягиваются друг к другу. В данной работе более подробно рассматриваются эти эффекты, а также строится обобщение результатов, полученных для случая чисто диссипативных связей~\cite{Pimenova-etal-2016}, на более общий случай глобальной связи типа Курамото-Сакагучи, которая может иметь как диссипативную, так и консервативную компоненты.

В качестве основной математической модели рассматривается ансамбль фазовых осцилляторов с глобальной связью типа Курамото-Сакагучи, на который действует общий шум:
\[
\dot{\varphi}_k=\varOmega_k +\frac{\mu}{N}\sum_{j=1}^N\sin(\varphi_j-\varphi_k-\beta) +\sigma\xi(t)\sin\varphi_k\,,\qquad
k=1,...,N\,.
\]
Здесь $\varOmega_k$ --- это собственная частота $k$-го осциллятора, $\sigma$ --- амплитуда общего шума, $\xi(t)$ --- нормированный гауссовский белый шум, $\mu$ --- коэффициент связи, $\beta$ характеризует фазовый сдвиг в члене связи (или, другими словами, вклады ``активной''  ($\sim\cos\beta$) и ``реактивной'' ($\sim\sin\beta$) компонент связи).
Восприимчивость фазы к шумовому воздействию, т.е.\ форма слагаемого $\sigma\xi(t)\sin{\varphi_k}$, соответствует случаю квазигармонических осцилляторов, подверженных действию общего линейно поляризованного шума, и некоторым другим физическим системам. Примером первого могут служить осцилляторы Ван-дер-Поля при слабой нелинейности и шуме в одной из переменных или метрономы, смонтированные на общей платформе, подверженной случайному силовому воздействию, как в экспериментальной работе~\cite{Martens-etal-2013}; примером второго являются связанные электрические осцилляторы~\cite{Temirbayev-etal-2012,Temirbayev-etal-2013}.
В термодинамическом пределе $N\gg1$ удобно параметризовать осцилляторы значением их собственной частоты $\varOmega$ и переписать уравнения динамики осцилляторов в виде
\begin{equation}
\dot\varphi_\varOmega=\varOmega +\mu R\sin(\varPhi-\varphi_\varOmega-\beta) +\sigma\xi(t)\sin\varphi_\varOmega\;,
\label{eq104}
\end{equation}
где введен параметр порядка $R$: $Re^{i\varPhi}\equiv\langle{e^{i\varphi_k}}\rangle$.
Форма уравнения~(\ref{eq104}), имеющего вид
\begin{equation}
\dot\varphi_\varOmega=\varOmega+\Im\left(H(t)e^{-i\varphi_\varOmega}\right)\;,
\label{eq101}
\end{equation}
где
\[
H(t)=\mu R e^{-i\beta} e^{i\varPhi}-\sigma\xi(t)\,,
\]
позволяет провести полный анализ коллективной динамики системы с помощью подходов Ватанабе-Строгаца и Отта-Антонсена~\cite{Watanabe-Strogatz-1994,Pikovsky-Rosenblum-2008,Marvel-Mirollo-Strogatz-2009,Ott-Antonsen-2008} и предоставляет возможность более глубокого изучения тонких аспектов взаимодействия между механизмами синхронизации/десинхронизации общим шумом и глобальной связью.
В термодинамическом пределе плотность распределения вероятности $w_\varOmega(\varphi,t)$ осцилляторов с собственными частотами  $\varOmega$ допускает решение, параметризованное одной комплексной величиной  $a_\varOmega$ (см.~\cite{Ott-Antonsen-2008} и Приложение),
\begin{equation}
w_\varOmega(\varphi,t)=\frac{1}{2\pi}\Big(
1+\sum_{j=1}^{\infty}[(a_\varOmega)^je^{ij\varphi}+c.c.]\Big)\,,
\label{eq102}
\end{equation}
где $a_\varOmega$ подчиняется уравнению
\begin{equation}
\dot{a}_\varOmega=-i\varOmega a_\varOmega+\frac{H^\ast(t)}{2}-\frac{H(t)}{2}a_\varOmega^2\,.
\label{eq103}
\end{equation}
Для плотности распределения фаз может быть вычислен комплексный параметр порядка (среднее поле):
\[
Re^{i\varPhi}=\int\limits_{-\infty}^{+\infty}\d\varOmega\,g(\varOmega)
\int\limits_{0}^{2\pi}\d\varphi\,e^{i\varphi}w_\varOmega(\varphi,t)=
\int\limits_{-\infty}^{+\infty}\d\varOmega\,g(\varOmega)\,a_\varOmega^\ast\,,
\]
где $g(\varOmega)$ --- плотность распределения собственных частот.

Главной задачей настоящей работы является обобщение результатов, полученных в~\cite{Pimenova-etal-2016}, где был рассмотрен случай $\beta=0$. Будет показано, что основными управляющими параметрами являются эффективная сила связи $\mu_\beta=\mu\cos\beta$ и эффективная частота $\varOmega_\beta=\varOmega-\mu\sin\beta$; часть полученных результатов будет аналогична результатам, полученным для $\beta=0$, с соответствующими эффективными параметрами.
Однако, динамика средней частоты осцилляторов оказывается существенно более сложной: в частности, в случае $\beta\ne0$ будет наблюдаться сдвиг средней частоты, в то время как для $\beta=0$ он исчезает.

Материал организован следующим образом. В разделе~\ref{sec2} рассмотрена динамика ансамбля абсолютно одинаковых осцилляторов, в этом случае возможно состояние полной синхронизации. Для одинаковых осцилляторов рассмотрены свойства устойчивости синхронного состояния и осредненная по времени динамика параметра порядка. Более тонкие характеристики, такие как плотность распределения вероятности для параметра порядка, могут быть аналитически найдены только в случае, если основная частота осцилляторов достаточно большая.

Раздел~\ref{sec3} посвящен рассмотрению более реалистичных систем осцилляторов, частоты которых отличаются друг от друга. В такой системе состояние полной синхронизации невозможно, но можно оценить среднюю по времени величину параметра порядка для состояния, близкого к синхронному, и состояния, близкого к абсолютно несинхронному. Более того, для случая высокочастотных осцилляторов оказывается возможным полностью описать динамику параметра порядка, в том числе вычислить плотность распределения вероятности.

В разделе ~\ref{sec4} найдены средние частоты осцилляторов для системы неодинаковых осцилляторов. Показано, что в таком ансамбле не происходит полного захвата частоты, как и в случае отсутствия общего шума.

\section{Ансамбль идентичных осцилляторов}\label{sec2}
В случае идентичных осцилляторов $\varOmega_k=\varOmega$, и, принимая во внимание то, что $a^*=R\exp(i\Phi)$, уравнение~\eqref{eq103} может быть переписано в виде
\begin{eqnarray}
&&\hspace{-15pt}
\dot{R}=\frac{\mu}{2}\cos\beta(1-R^2)R-\frac{\sigma\xi(t)}{2}(1-R^2)\cos\varPhi\,,
\label{eq202}\\[10pt]
&&\hspace{-15pt}
\dot\varPhi=\varOmega-\frac{\mu}{2}\sin\beta(1+R^2)
+\frac{\sigma\xi(t)}{2}\left(\frac{1}{R}+R\right)\sin\varPhi\,.
\label{eq203}
\end{eqnarray}
Для удобства введем новый параметр порядка $J=R^2/(1-R^2)$ (то есть $R=\sqrt{J/(1+J)}$). В новых переменных уравнения в смысле Стратоновича принимают следующий вид:
\begin{align}
\dot{J}&=\mu\cos\beta\,J-\sigma\xi(t)\sqrt{J(1+J)}\cos\varPhi\,,
\label{eq204}\\
\dot\varPhi&=\varOmega-\mu\sin\beta\frac{J+1/2}{J+1} +\sigma\xi(t)\frac{J+1/2}{\sqrt{J(1+J)}}\sin\varPhi\,.
\label{eq205}
\end{align}

В терминах $J$ состоянию полной синхронизации ($R\to 1$) соответствует $J\to\infty$, а состоянию максимальной асинхронности ($R=0$) соответствует $J=0$. Полученная система уравнений может быть исследована аналитически для $J\gg1$ (приближение к полной синхронизации) и $J\ll1$ (приближение к максимальной асинхронности).

\subsection{Устойчивость синхронного состояния: $J\gg1$}
Для $J\gg1$ в ведущем порядке система уравнений~\eqref{eq204}--\eqref{eq205} имеет вид
\begin{eqnarray}
&&\dot{J}=\mu_\beta J-\sigma\xi(t)J\cos\varPhi\,,
\label{eq2a01}\\[10pt]
&&\dot\varPhi=\varOmega_\mu+\sigma\xi(t)\sin\varPhi\,,
\label{eq2a02}
\end{eqnarray}
где $\mu_\beta\equiv\mu\cos\beta$ и $\varOmega_\mu\equiv\varOmega-\mu\sin\beta$.
Уравнение~(\ref{eq2a01}) может быть переписано в виде
\[
\frac{\d}{\d t}\ln J=\mu_\beta-\sigma\xi(t)\cos\varPhi\,,
\]
откуда видно, что показатель Ляпунова $\lambda\equiv\big\la\frac{\d}{\d t}\ln J\big\ra$ оказывается
\begin{equation}
\lambda=\mu_\beta-\sigma\la\xi(t)\cos\varPhi\ra
=\mu_\beta+\sigma^2\la\sin^2\varPhi\ra\,,
\label{eq2a03}
\end{equation}
где $\langle\cdot\rangle$ означает осреднение по реализациям шума. Положительное значение $\lambda$ означает, что состояние синхронизации $J=\infty$ является устойчивым. Видно, что шум вносит положительный вклад в $\lambda$, тогда как связь может вносить как положительный, так и отрицательный вклад.

Динамика $\varPhi$ управляется уравнением~\eqref{eq2a02} и не зависит от $J$. На основании этого можно написать уравнение Фоккера--Планка для плотности вероятности $W(\varPhi,t)$:
\begin{equation}
\frac{\partial}{\partial t}W
+\frac{\partial}{\partial\varPhi}(\varOmega_\mu W)
-\sigma^2\frac{\partial}{\partial\varPhi}\left(\sin\varPhi
\frac{\partial}{\partial\varPhi}\big(\sin\varPhi\,W
\big)
\right)=0\,.
\label{eq2a04}
\end{equation}
Это уравнение допускает стационарное $\pi$-периодичное решение
\[
W(\varPhi)=\frac{C}{\sin\varPhi}
\int\limits_\varPhi^\pi\frac{\d\varPhi_1}{\sin\varPhi_1}e^{\frac{\varOmega_\mu}{\sigma^2}(\ctg\varPhi_1-\ctg\varPhi)},
\]
где константа $C$ определяется из условия нормировки $\int_0^{2\pi}W(\varPhi)\,\d\varPhi=1$.

Тогда
\begin{equation}
\la\sin^2\varPhi\ra=\frac
{\displaystyle
	\int\limits_0^\pi\d\varPhi\,\sin\varPhi
	\int\limits_\varPhi^\pi\frac{\d\varPhi_1}{\sin\varPhi_1}e^{\frac{\varOmega_\mu}{\sigma^2}(\ctg\varPhi_1-\ctg\varPhi)}}
{\displaystyle
	\int\limits_0^\pi\frac{\d\varPhi}{\sin\varPhi}
	\int\limits_\varPhi^\pi\frac{\d\varPhi_1}{\sin\varPhi_1}e^{\frac{\varOmega_\mu}{\sigma^2}(\ctg\varPhi_1-\ctg\varPhi)}}\,.
\label{eq2a05}
\end{equation}

\begin{figure}[!t]
	\centerline{
		\includegraphics[width=0.64\textwidth]%
		{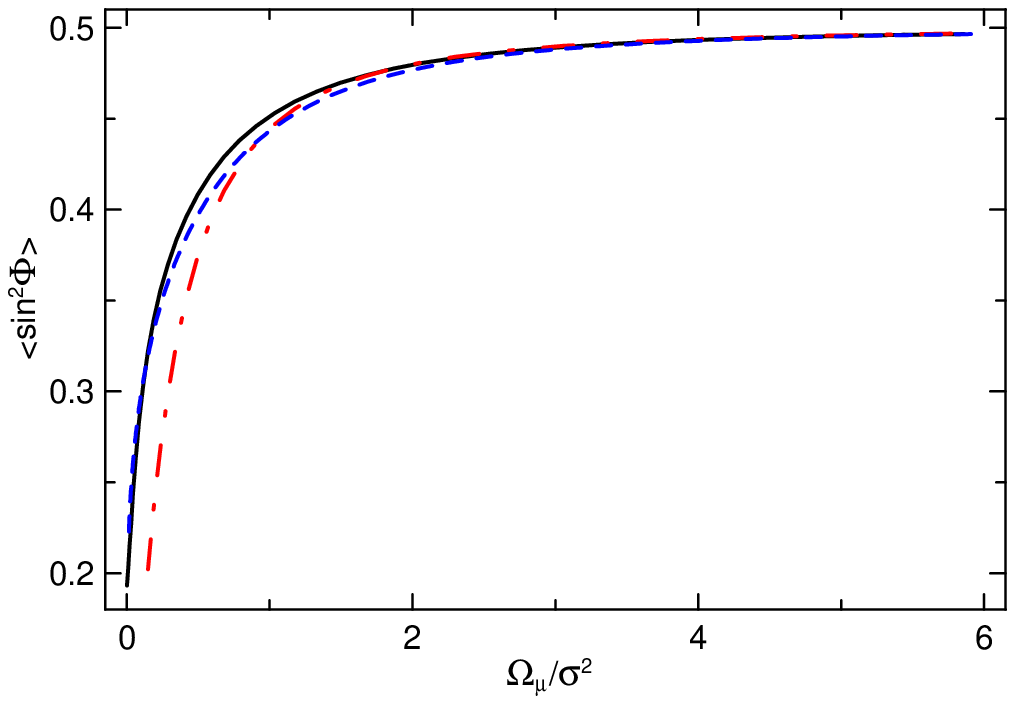}
	}
	
	\caption{
Зависимость $\la\sin^2\varPhi\ra$ от $\varOmega_\mu/\sigma^2$ определяет показатель Ляпунова $\lambda$ (см.\ уравнение~(\ref{eq2a03})). Сплошной линией построено точное решение (\ref{eq2a05}), пунктирная линия соответствует аппроксимации Галеркина (\ref{eq2gal3})--(\ref{eq2gal4}), штрихпунктирная линия соответствует асимптотическому разложению (\ref{eq2a06}). Для точного решения и аппроксимации Галеркина $\la\sin^2\varPhi\ra$ стремится к ненулевому конечному значению (примерно $0.2$) при $\varOmega_\mu/\sigma^2\to0$.
	}
	\label{fig1}
\end{figure}

Зависимость $\la\sin^2\varPhi\ra$ от $\varOmega_\mu/\sigma^2$ показана на рис.~\ref{fig1}.
Видно, что влияние общего шума на устойчивость состояния синхронизации гораздо сильнее выражено для высокочастотных колебаний, и это влияние монотонно уменьшается с уменьшением собственной частоты осцилляторов. Ниже приводятся два способа, позволяющие вычислить показатель Ляпунова для высокочастотных осцилляторов без вычисления интегралов в~\eqref{eq2a05}.

\paragraph{Случай малого шума $\sigma^2\ll\varOmega_\mu$.}
Интеграл в уравнении~(\ref{eq2a05}) в общем случае не может быть вычислен аналитически. Однако, можно оценить его асимптотическое поведение при $\varOmega_\mu\gg\sigma^2$. Принимая $\sigma^2$ за малый параметр в уравнении~(\ref{eq2a04}), можно представить функцию  плотности распределения вероятности $W(\varPhi)$ в виде ряда $W(\varPhi)=(2\pi)^{-1}+\sigma^2W^{(1)}(\varPhi)+\sigma^4W^{(2)}(\varPhi)+\dots$, и найти
$\la\sin^2\varPhi\ra=\frac12-\frac{\sigma^4}{8\varOmega_\mu^2}+\dots$.
Тогда
\begin{equation}
\lambda=\mu_\beta+\sigma^2\left[\frac12-\frac{\sigma^4}{8\varOmega_\mu^2}+\mathcal{O}\left(\frac{\sigma^6}{\varOmega_\mu^3}\right)\right]\,.
\label{eq2a06}
\end{equation}

\paragraph{Приближение Галеркина.}
Двухпараметрическая функция
\begin{equation}
W_{m,\varPhi_0}(\varPhi)=\frac{a_m}{\sqrt{1+m\sin^2(\varPhi-\varPhi_0)}}\,,
\label{eq2gal1}
\end{equation}
где $a_m=\big(4\mathrm{K}\sqrt{-m}\big)^{-1}$,
позволяет достаточно точно аппроксимировать функцию $W(\varPhi)$, определяемую  уравнением Фоккера--Планка~(\ref{eq2a04}), и может быть использована в качестве аппроксимирующей функции для метода Галеркина~\cite{Fletcher-Galerkin_method}.
Здесь и далее $\mathrm{K}(z)$ и $\mathrm{E}(z)$ обозначают эллиптические интегралы первого и второго рода соответственно. Заметим, что эти интегралы являются действительными конечными функциями чисто мнимого аргумента на $[0,+i\infty)$.
Выполним процедуру проецирования уравнения Фоккера--Планка~(\ref{eq2a04}) на подпространство функций $W_{m,\varPhi_0}$ (см.\ уравнение~(\ref{eq2gal1})). В случае стационарного состояния, уравнение~(\ref{eq2a04}) может быть проинтегрировано:
\begin{equation}
\frac{\varOmega_\mu}{\sigma^2}W-\sin\varPhi
\frac{\partial}{\partial\varPhi}\big(\sin\varPhi\,W
\big)=\frac{j}{\sigma^2}\,
\label{eq2gal2}
\end{equation}
где $j$  --- это постоянный по $\varPhi$ поток вероятности. При $\varOmega_\mu/\sigma^2\to0$,
решение уравнения~(\ref{eq2gal2}) имеет вид
$\lim_{m\to\infty,\varPhi_0\to0}W_{m,\varPhi_0}(\varPhi)$.
При $\varOmega_\mu/\sigma^2\to\infty$, вид решения уравнения ~(\ref{eq2gal2}) полностью соответствует $W_{m,\varPhi_0}(\varPhi)$ с $m\ll1$.

Перепишем уравнение~(\ref{eq2gal2}) в операторном виде $\hat{\mathcal{L}}W-j/\sigma^2=0$. Для минимизации невязки аппроксимации можно потребовать выполнения следующих условий:
\[
\int\limits_0^{2\pi}\d\varPhi(\hat{\mathcal{L}}W_{m,\varPhi_0}-j/\sigma^2)=
\int\limits_0^{2\pi}\d\varPhi\,W_{m,\varPhi_0}(\hat{\mathcal{L}}W_{m,\varPhi_0}-j/\sigma^2)=
\int\limits_0^{2\pi}\d\varPhi\,\frac{\d W_{m,\varPhi_0}}{\d\varPhi}(\hat{\mathcal{L}}W_{m,\varPhi_0}-j/\sigma^2)=0\,,
\]
которые позволяют определить константу $j$ и параметры $m$ и $\varPhi_0$:
\[
\sin^2\varPhi_0=\frac{1}{2+m}\,,\quad j=\frac{2\pi a_m^2\varOmega_\mu}{\sqrt{1+m}}\,,
\]
\begin{equation}
\frac{\varOmega_\mu}{\sigma^2}=\frac{\sqrt{1+m}}{m}
\frac{\displaystyle
	\frac{2}{2+m}\frac{\mathrm{E}(\sqrt{-m})}{\mathrm{K}(\sqrt{-m})}-1
}{\displaystyle
	1-\frac{1}{\sqrt{1+m}}
	\left(\frac{\pi}{2\mathrm{K}(\sqrt{-m})}\right)^2
}\,,
\label{eq2gal3}
\end{equation}
и
\begin{equation}
\la\sin^2\varPhi\ra=\frac{1}{2+m}\frac{\mathrm{E}(\sqrt{-m})}{\mathrm{K}(\sqrt{-m})}\,.
\label{eq2gal4}
\end{equation}
С учетом уравнений~(\ref{eq2gal3}) и (\ref{eq2gal4}), можно найти зависимость $\la\sin^2\varPhi\ra$ и $\varOmega_\mu/\sigma^2$, параметризованную $m$.

На рис.~\ref{fig1} видно, что аппроксимация Галеркина достаточно хорошо согласуется с точным решением~(\ref{eq2a05}), и оценить область применимости асимптотической формулы~(\ref{eq2a06}).

\subsection{Динамика $\la{J}\ra_\xi(t)$ и потеря максимальной асинхронности}
Уравнение динамики осредненного параметра порядка $\la{J}\ra_\xi$ (где $\la...\ra_\xi$ обозначает осреднение по реализациям шума) наиболее просто получить из системы уравнений~(\ref{eq204})--(\ref{eq205}), записанных в форме Ито:
\begin{eqnarray}
&&\hspace{-15pt}
\dot{J}=\mu_\beta J+\sigma^2\left(J+\frac12\right)-\sigma\xi(t)\circ\sqrt{J(1+J)}\cos\varPhi,
\label{eq2bJ1}\\[10pt]
&&\hspace{-15pt}
\dot\varPhi=\varOmega-\mu\sin\beta\frac{J+1/2}{J+1}
 +\sigma^2\left(\frac32+\frac12\frac{1}{J(1+J)}\right)\sin\varPhi\cos\varPhi+{}
\nonumber\\[7pt]
&&\qquad\qquad\qquad
 {}+\sigma\xi(t)\circ\frac{J+1/2}{\sqrt{J(1+J)}}\sin\varPhi.
\label{eq2bJ2}
\end{eqnarray}
Так как в форме Ито мгновенное состояние системы не зависит от шумового сигнала в этот же момент времени,  $\la\xi(t)\circ f(J(t),\varPhi(t))\ra_\xi=0$, а детерминированная часть уравнения~(\ref{eq2bJ1}) содержит только константы и линейные по $J$ члены, можно осреднить уравнение~(\ref{eq2bJ1}) по реализациям шума и получить
\begin{equation}
\frac{\d}{\d t}\la{J}\ra_\xi=(\mu_\beta+\sigma^2)\la{J}\ra_\xi+\frac{\sigma^2}{2}\,.
\label{eq2bJ3}
\end{equation}
Решение уравнения~(\ref{eq2bJ3}), соответствующее тому, что система изначально находится в состоянии максимальной асинхронности ($J(0)=0$), имеет вид:
\begin{equation}
\la{J}\ra_\xi(t)=(e^{(\mu_\beta+\sigma^2)t}-1)\frac{\sigma^2}{2(\mu_\beta+\sigma^2)}\,.
\label{eq2bJ4}
\end{equation}

Условие роста среднего значения параметра порядка $\la{J}\ra_\xi$
\begin{equation}
\mu_\beta+\sigma^2>0
\label{cond1}
\end{equation}
не совпадает с условием положительности показателей Ляпунова
\begin{equation}
\mu_\beta+\sigma^2\la\sin^2\varPhi\ra>0\;,
\label{cond2}
\end{equation}
так как в уравнении~\eqref{cond1} вклад больших значений $J$ является более существенным.
Уравнение~(\ref{cond2}) определяет, стремится ли система асимптотически к состоянию полной синхронизации $J=\infty$, тогда как уравнение~(\ref{cond1}) определяет, стремится ли система уйти от состояний максимальной асинхронности, но при этом не запрещает системе возвращаться в слабо синхронные состояния от состояний с большими значениями $J$.
Таким образом, условие~(\ref{cond1}) слабее, чем~(\ref{cond2}).

\subsection{Переход от максимальной асинхронности к синхронизации}
Между состояниями полной синхронизации ($J=\infty$) и максимальной асинхронности ($J=0$) существует одно значительное отличие. Синхронное состояние является притягивающим, $\lambda=\mu_\beta+\sigma^2\la\sin^2\varPhi\ra>0$,
переход в это состояние необратим, тогда как состояние десихронизации  $\mu_\beta+\sigma^2<0$ не притягивает к себе фазовые траектории. Более того, так как шумовое слагаемое в~\eqref{eq202} не исчезает при $R=0$, шум ``выбрасывает'' систему из этого состояния. Таким образом, так как синхронное состояние является притягивающим, переход к синхронизации является однонаправленным, и интерес представляет время перехода, которое может быть найдено из уравнения Фоккера--Планка для плотности распределения вероятности $W(J,\varPhi,t)$.
Для системы стохастических уравнений (\ref{eq204})--(\ref{eq205}) оно имеет вид
\begin{equation}
\frac{\partial}{\partial t}W+\frac{\partial}{\partial J}\left(\mu_\beta J\,W\right)
+\frac{\partial}{\partial\varPhi}\left(\left[\varOmega
-\mu\sin\beta\frac{J+1/2}{J+1}\right]\,W\right)
-\sigma^2\hat{Q}^2W=0\,,
\label{eq2c01}
\end{equation}
где оператор $\hat{Q}$ определен как
\[
\hat{Q}(\cdot)\equiv\frac{\partial}{\partial J}
\bigg(-\sqrt{J(1+J)}\cos\varPhi\,(\cdot)\bigg)
+\frac{\partial}{\partial\varPhi}
\bigg(\frac{J+1/2}{\sqrt{J(1+J)}}\sin\varPhi\,(\cdot)\bigg)\,.
\]
Это уравнение может быть исследовано аналитически для физически реалистичного случая больших частот $\varOmega\gg\mu\sim\sigma^2$.

\subsubsection{Осреднение по высокочастотным колебаниям}
Можно показать, что при исчезающе малых $\mu$ и $\sigma$ распределение плотности вероятности принимает вид $W(J,\varPhi,t)=(2\pi)^{-1}w(J,t)$, где $\int_0^\infty w(J,t)\,\d J=1$. Тогда для $\mu\sim\sigma^2\ll\varOmega$ можно считать, что $\mu=\sigma^2\mu_1$, $\varOmega=\varOmega_0$ и воспользоваться методом многих масштабов: $W=W^{(0)}(J,t_1,t_2,...)+\sigma^2W^{(1)}(J,\varPhi,t_0,t_1,t_2...)+...$\,, где $t_n=\sigma^{2n}t$. В ведущем порядке уравнение~(\ref{eq2c01}) дает $W^{(0)}=(2\pi)^{-1}w(J,t_1,t_2,...)$. В следующем порядке, $\sigma^2$, уравнение~(\ref{eq2c01}) дает
\begin{eqnarray}
&&\hspace{-15pt}
\frac{\partial W^{(1)}}{\partial t_0}+\varOmega_0\frac{\partial W^{(1)}}{\partial\varPhi}
+\frac{\partial W^{(0)}}{\partial t_1}
+\frac{\partial}{\partial J}\left(\mu_1\cos\beta\,J\,W^{(0)}\right)+{}
\nonumber\\[7pt]
&&
{}
+\frac{\partial}{\partial\varPhi}\left(-\mu_1\sin\beta\frac{J+1/2}{J+1}\,W^{(0)}\right)
-\hat{Q}^2W^{(0)}=0\,.
\nonumber
\end{eqnarray}
Проинтегрировав последнее уравнение по $\varPhi$ от $0$ до $2\pi$, находим
\[
\frac{\partial}{\partial t_0}\int\limits_0^{2\pi}W^{(1)}\d\varPhi
+\frac{\partial w(J,t_1)}{\partial t_1}
+\frac{\partial}{\partial J}\left(\mu_1\cos\beta\,J\,w(J,t_1)\right)
-\frac{1}{2\pi}\int\limits_0^{2\pi}\hat{Q}^2w(J,t_1)\,\d\varPhi=0\,.
\]
Чтобы избежать линейного роста $W^{(1)}$ по $t_0$, необходимо обратить в ноль первое слагаемое. Вычисляя последний интеграл
\[
\frac{1}{2\pi}\int_0^{2\pi}\hat{Q}^2w(J,t_1)\,\d\varPhi
=\frac{1}{2}\frac{\partial}{\partial J}\left(
\sqrt{J(1+J)}\left[\frac{\partial}{\partial J}\left(\sqrt{J(1+J)}\,w\right)
-\frac{J+1/2}{\sqrt{J(1+J)}}w\right]\right)\,,
\]
получим, что эволюция плотности распределения вероятности $w(J,t)$ определяется уравнением
\begin{equation}
\frac{\partial w}{\partial t}
+\frac{\partial}{\partial J}\left(\left[\mu_\beta J
+\frac{\sigma^2}{2}\left(J+\frac12\right)\right]w\right)
-\sigma^2\hat{Q}_J^2w=0\,,
\label{eq2c02}
\end{equation}
где
\[
\hat{Q}_J(\cdot)\equiv\frac{1}{\sqrt{2}}\frac{\partial}{\partial J}\Big(\sqrt{J(1+J)}\,(\cdot)\Big)\,.
\]
Уравнение~(\ref{eq2c02}) может быть интерпретировано как уравнение Фоккера--Планка для стохастического уравнения
\begin{equation}
\dot{J}=\mu_\beta J+\frac{\sigma^2}{2}(J+1/2)
+\sigma\sqrt{\frac{J(1+J)}{2}}\,\zeta(t)\,,
\label{eq2c03}
\end{equation}
где эффективный шум $\zeta(t)$ является гауссовым и дельта-коррелированным, $\la\zeta(t)\zeta(t+t')\ra=2\delta(t')$\,.



\paragraph{Сильная десинхронизирующая связь ($\mu_\beta<-\sigma^2/2$).}
В этом случае уравнение (\ref{eq2c02}) допускает стационарное решение с нулевым потоком вероятности:
\begin{equation}
w_0(J)=\left(\frac{2(-\mu_\beta)}{\sigma^2}-1\right)(1+J)^{-\frac{2(-\mu_\beta)}{\sigma^2}}\,.
\label{eq2c05}
\end{equation}
Это означает, что в системе нет притягивающих состояний. Синхронное состояние не является асимптотически притягивающим в этом случае ($\lambda<0$).
Плотность вероятности~(\ref{eq2c05}) позволяет найти среднее значение и дисперсию среднего поля
\[
\la{R}\ra=\lla\sqrt{\frac{J}{1+J}}\rra
=\frac{\sqrt{\pi}}{2}\frac{\Gamma\big(2(-\mu_\beta)/\sigma^2\big)}{\Gamma\big(2(-\mu_\beta)/\sigma^2+1/2\big)}\,,
\qquad
\la{R^2}\ra=\lla\frac{J}{1+J}\rra=\frac{\sigma^2}{2(-\mu_\beta)}\,.
\]

\paragraph{Асимптотически притягивающее синхронное состояние ($\mu_\beta>-\sigma^2/2$).}
В этом случае формальное решение может быть написано только для конечного потока вероятности  $j$\,:
\begin{equation}
w_j(J)=\frac{2j}{\sigma^2}(1+J)^{2\mu_\beta/\sigma^2}
\int_J^\infty\frac{\d J_1}{J_1}(1+J)^{-1-2\mu_\beta/\sigma^2}.
\label{eq2c06}
\end{equation}
Распределение плотности вероятности $w_j(J)|_{J\gg1}\propto 1/J_1$ имеет ``тяжелые'' хвосты, интегралы от которых расходятся.
После нормировки можно найти $w(J<\infty)=0$, тогда как $\int_J^{\infty}w_1(J_1)\,\d J_1=1$, что соответствует $j\to0$ и также означает, что все состояния ``собираются'' при $J=\infty$.

\subsubsection{Переход к синхронному состоянию: Время перехода}
В случае, когда состояние полной синхронизации является притягивающим, интерес представляет нахождение характерного времени перехода из состояния максимальной асинхронности к синхронному состоянию. Строго говоря, так как система может достигнуть состояния полной синхронизации только асимптотически, время перехода к этому состоянию всегда бесконечно. Однако, можно рассмотреть, как система приближается к состоянию, близкому к полной синхронности, и найти время перехода к некоторому большому значению $\bar J$. Для уравнения~(\ref{eq2c03}) (или уравнения~(\ref{eq2c02})), время перехода $T(J_0,\bar J)$ от $J_0$ к $\bar J$ определяется уравнением
\begin{equation}
B(J_0)\frac{\partial^2T(J_0,\bar J)}{\partial J_0^2}+A(J_0)\frac{\partial T(J_0,\bar J)}{\partial J_0}=-1\,,
\label{eq2fpt1}
\end{equation}
где $T(\bar J,\bar J)=0$, $\big(\partial T(J_0,\bar J)/\partial J_0\big)|_{J_0=0}=0$ (эффективное отражающее граничное условие в $J_0=0$) и
$A(J_0)=\mu_\beta J_0+\sigma^2(J_0+1/2)$,
$B(J_0)=\frac{\sigma^2}{2}J_0(1+J_0)$.
Заметим, что в рассматриваемой задаче $J_0=0$ является границей области возможных состояний системы, с чем и связано эффективное граничное условие в этой точке. Решение уравнения~(\ref{eq2fpt1}) имеет вид
\[
T(J_0,\bar J)=\int_{J_0}^{\bar J}\d J_1\int_{0}^{J_1}\frac{\d J_2}{B(J_2)}
e^{-\int\limits_{J_2}^{J_1}\frac{A(J_3)}{B(J_3)}\d J_3}
=\frac{2}{\sigma^2}\int_{J_0}^{\bar J}\frac{\d J_1}{J_1}
\int_{0}^{J_1}\d J_2\frac{(1+J_2)^{2\mu_\beta\sigma^{-2}}}{(1+J_1)^{2\mu_\beta\sigma^{-2}+1}}\,.
\]
Интегрируя по $J_2$ и полагая $J_0\to0$, можно найти
\begin{eqnarray}
T(0,\bar J)=\frac{\sigma^{-2}}{\mu_\beta\sigma^{-2}+1/2}
\int_0^{\bar J}\!\frac{\d z}{z}\!
\left(1-\frac{1}{(1+z)^{2\mu_\beta\sigma^{-2}+1}}\right)={}
\qquad\quad\nonumber \\[10pt]
=\frac{1}{\sigma^2}\left[\frac{\ln(1+\bar J)}{\mu_\beta\sigma^{-2}+1/2}
+\frac{\bar J^{-(2\mu_\beta\sigma^{-2}+1)}}{2(\mu_\beta\sigma^{-2}+1/2)^2}
+\tau\left(\frac{2\mu_\beta}{\sigma^2},\bar J\right)\right],
\label{eq3fpt2}
\end{eqnarray}
где $\tau(q,\bar J)$ мало по сравнению с суммой первого и второго слагаемого в скобках  при $\bar J\gg1$.
При $2\mu_\beta\sigma^{-2}+1>0$ время перехода является логарифмически большим $\sim\ln{\bar J}$, и это означает, что синхронное состояние притягивает траектории. При $2\mu_\beta\sigma^{-2}+1<0$, время перехода имеет степенную зависимость от $\bar J$, то есть в этом случае синхронное состояние является отталкивающим, и фазовые траектории системы редко проходят вблизи этого состояния.

\paragraph{Физическая интерпретация для идентичных осцилляторов.} Результаты данного раздела могут быть резюмированы в виде следующей качественной картины. Для конкуренции между воздействиями связи и общего шума имеется критическое значение параметра связи $\mu_\beta^\mathrm{crit}=-\sigma^2/2$. При $\mu_\beta>\mu_\beta^\mathrm{crit}$ превалирует воздействие шума: со временем ансамбль асимптотически приближается к состоянию полной синхронизации. При $\mu_\beta<\mu_\beta^\mathrm{crit}$ отталкивающая связь предотвращает полную синхронизацию. Однако, в последнем случае параметр порядка никогда не стремиться к нулю: наблюдается частичная синхронизация с флуктуирующим параметром порядка.

\section{Неидентичные осцилляторы:\\ Переход к синхронизации}\label{sec3}
Рассмотрим ансамбль осцилляторов, имеющих разные собственные частоты $\varOmega$. Будем считать, что частоты имеют Лоренцевское распределение с характерной шириной $\gamma$:
\[
g(\varOmega)=\frac{\gamma}{\pi[\gamma^2+(\varOmega-\varOmega_0)^2]}\,.
\]
Функция $a(\varOmega)$ может быть рассмотрена как аналитическая функция комплексного аргумента $\varOmega$. 
Тогда интеграл с распределением $g(\varOmega)$ может быть вычислен методами теории вычетов:
\[
R\exp[-i\Phi]=\int\limits_{-\infty}^{+\infty}\d\varOmega\,g(\varOmega)\,a(\varOmega)=a(\varOmega_0-i\gamma)\,.
\]
Уравнение~(\ref{eq103}), записанное для $a(\varOmega_0-i\gamma)$, дает замкнутое уравнение для параметра порядка, которое в терминах вещественных переменных имеет вид
\begin{eqnarray}
&&\dot{R}=-\gamma R+\frac{\mu_\beta}{2}(1-R^2)R-\frac{\sigma\xi(t)}{2}(1-R^2)\cos\varPhi ,
\label{eq303}\\[10pt]
&&\dot\varPhi=\varOmega_0
-\frac{\mu}{2}\sin\beta(1+R^2)
+\frac{\sigma\xi(t)}{2}\!\left(\!\frac{1}{R}+R\!\right)\sin\varPhi .
\label{eq304}
\end{eqnarray}
В терминах $J$ и $\varPhi$, полученная система уравнений имеет вид
\begin{eqnarray}
&&\dot{J}=\mu_\beta J-2\gamma J(1+J)-\sigma\xi(t)\sqrt{J(1+J)}\cos\varPhi,
\label{eq305}\\[10pt]
&&\dot\varPhi=\varOmega_0
-\mu\sin\beta\frac{J+1/2}{J+1}
+\sigma\xi(t)\frac{J+1/2}{\sqrt{J(1+J)}}\sin\varPhi.
\label{eq306}
\end{eqnarray}

\subsection{Состояния, близкие к синхронным ($J\gg1$)}
При $J\gg1$ уравнения~(\ref{eq305})--(\ref{eq306}) имеют вид
\begin{eqnarray}
&&\hspace{-15pt}
\dot{J}=\mu_\beta J -2\gamma J^2-\sigma\xi(t)J\cos\varPhi\,,
\label{eq3a01}\\[10pt]
&&\hspace{-15pt}
\dot\varPhi=\varOmega_{\mu,0}+\sigma\xi(t)\sin\varPhi\,.
\label{eq3a02}
\end{eqnarray}
Аналогично случаю уравнений~(\ref{eq2a01})--(\ref{eq2a02}), можно найти
\[
\la\frac{\d}{\d t}\ln{J}\ra=\mu_\beta-2\gamma\la{J}\ra -\sigma\la\xi(t)\cos\varPhi\ra
=\mu_\beta+\sigma^2\la\sin^2\varPhi\ra-2\gamma\la{J}\ra\,,
\]
где $\la\sin^2\varPhi\ra$ определяется уравнением~(\ref{eq2a05}). В случае неидентичных осцилляторов $\gamma\ne0$, и система не может достигнуть состояния полной синхронизации. В стационарном случае среднее значение производной по времени от $\ln{J}$ обращается в ноль. Тогда
\begin{equation}
\la{J}\ra\approx\frac{\lambda}{2\gamma}\,.
\label{eq3a03}
\end{equation}
Последнее уравнение справедливо при $\la{J}\ra\gg1$, что наблюдается при малой расстройке частот $\gamma\ll|\lambda|$.

\subsection{Потеря максимальной асинхронности}
Записывая уравнения~\eqref{eq305} и \eqref{eq306} в форме Ито и осредняя их по реализациям шума подобно тому, как это было сделано для вывода уравнения~\eqref{eq2bJ3} в случае идентичных осцилляторов, можно получить
\begin{equation}
\frac{\d}{\d t}\la{J}\ra_\xi =(\mu_\beta+\sigma^2-2\gamma)\la{J}\ra_\xi-2\gamma\la{J^2}\ra_\xi+\frac{\sigma^2}{2}\,.
\label{eq3b01}
\end{equation}
Для системы, изначально находящейся в максимально асинхронном состоянии ($J(0)=0$), до тех пор, пока $J\ll 1$, уравнение~(\ref{eq3b01}) дает решение вида
\begin{equation}
\la{J}\ra_\xi(t)\approx(e^{(\mu_\beta+\sigma^2-2\gamma)t}-1)\frac{\sigma^2}{2(\mu_\beta+\sigma^2-2\gamma)}\,.
\label{eq3b02}
\end{equation}
Так же, как и для случая идентичных осцилляторов, дальнейший анализ переходного поведения системы в общем случае не представляется возможным. Более подробный анализ возможен для осцилляторов с высокой собственной частотой $\varOmega\gg\mu\sim\sigma^2$.

\subsection{Высокочастотные осцилляторы}
Осреднение по высокочастотным колебаниям может быть выполнено точно так же, как и для случая одинаковых осцилляторов.
В этом случае уравнение~(\ref{eq2c03}) примет вид
\begin{equation}
\dot{J}=\mu_\beta J-2\gamma J(1+J)
+\frac{\sigma^2}{2}(J+1/2)+\sigma\sqrt{\frac{J(1+J)}{2}}\,\zeta(t)\,.
\label{eq3c01}
\end{equation}
Для ненулевого $\gamma$ состояние полной синхронности оказывается невозможным, поэтому параметр порядка всегда находится в пределах $0\leq J<\infty$. Проинтегрировав уравнение Фоккера--Планка для стационарной плотности вероятности $w(J)$, находим
\begin{equation}
w(J)=\frac{ \displaystyle
	(1+J)^{2\mu_\beta/\sigma^2}
	\exp\!\left[-\frac{4\gamma}{\sigma^2}(1+J)\right]}
{  \displaystyle
	\left(\frac{\sigma^2}{4\gamma}\right)^{2\mu_\beta/\sigma^2+1}
	\Gamma\!\left(\frac{2\mu_\beta}{\sigma^2}+1,\frac{4\gamma}{\sigma^2}\right)}\,,
\label{eq3c03}
\end{equation}
где $\Gamma(m,x)$ --- это верхняя неполная Гамма-функция.
Полученное решение позволяет выразить следующие моменты параметров порядка:
\begin{equation}
\la{R^2}\ra=1-\frac{4\gamma}{\sigma^2}\frac{
	\displaystyle
	\Gamma\!\left(\frac{2\mu_\beta}{\sigma^2},\frac{4\gamma}{\sigma^2}\right)}
{ \displaystyle
	\Gamma\!\left(\frac{2\mu_\beta}{\sigma^2}+1,\frac{4\gamma}{\sigma^2}\right)}\,,\quad
\la{J}\ra=
\frac{\sigma^2}{4\gamma}\frac{
	\displaystyle
	\Gamma\!\left(\frac{2\mu_\beta}{\sigma^2}+2,\frac{4\gamma}{\sigma^2}\right)}
{ \displaystyle \Gamma\!\left(\frac{2\mu_\beta}{\sigma^2}+1,\frac{4\gamma}{\sigma^2}\right)}-1\,.
\label{eq3c05}
\end{equation}

\begin{figure}[!t]
	\centerline{
		\includegraphics[width=0.60\textwidth]%
		{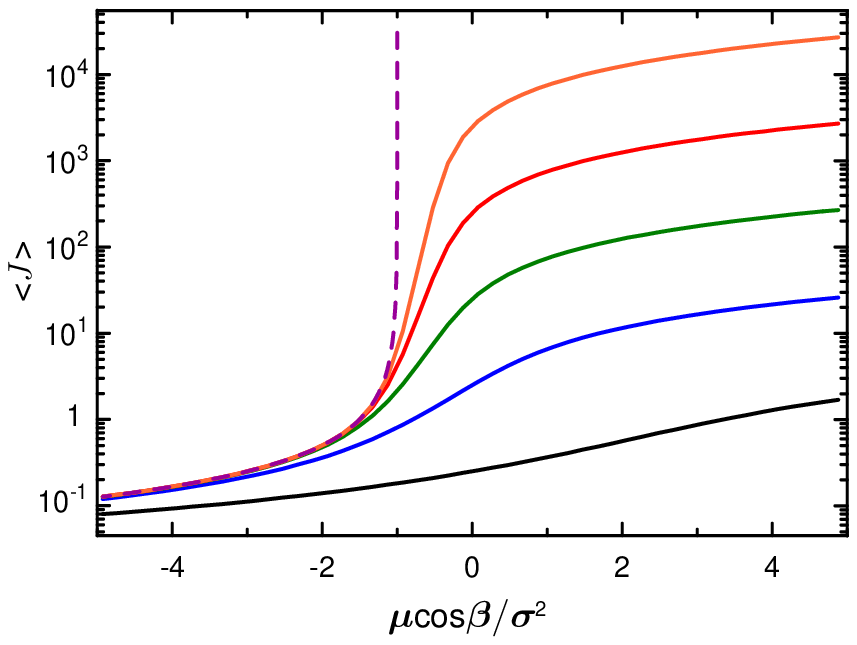}
	}
	
	\caption{
Сплошными линиями показана зависимость параметра порядка $\la{J}\ra$ от $\mu\cos\beta/\sigma^2$ ~(см.\ уравнене~\eqref{eq3c05}) при различных значениях $\gamma=1$, $10^{-1}$, $10^{-2}$, $10^{-3}$, $10^{-4}$ (снизу вверх соответственно), пунктирная линия, стремящаяся к бесконечности  при $\mu\cos\beta/\sigma^2=-1$, соответствует $\gamma=0$.
	}
	\label{fig2}
\end{figure}

На рис.~\ref{fig2} построено среднее значение параметра порядка $\la{J}\ra$ в зависимости от силы связи $\mu_\beta$ при различных значениях $\gamma$. Примечательно, что для иднтичных осцилляторов состояние полной синхронности является притягивающим при $\mu\cos\beta/\sigma^2>-1/2$.

\paragraph{Физическая интерпретация для коллективной динамики неидентичных осцилляторов.} Ансамбль неидентичных осцилляторов под действием общего шума ни синхронизируется, ни десинхронизируется. Для любой силы связи и уровня шума параметр порядка флуктуирует в диапазоне $0\le R<1$.

\section{Неидентичные осцилляторы:\\ Взаимное притяжение и отталкивание частот}\label{sec4}
Обращаясь к вопросу о поведении средних частот неидентичных связанных осцилляторов, следует сразу отметить одно существенное отличие между случаями наличия и отсутствия шума. В отсутствие шума достаточно сильная притягивающая связь может полностью синхронизовать частоты двух неидентичных осцилляторов; однако, при наличии шума идеального совпадения средних частот не наблюдается даже при сколь угодно большой силе связи (см.\ вставки (a) и (b) на рис.\ 3 в \cite{Pimenova-etal-2016}, где приведены результаты численного счета для ансамбля Курамото, $\beta=0$). Это обусловлено тем, что для случайного сигнала не запрещены периоды аномально больших возмущений, при которых сила притягивающей связи может становиться временно недостаточной для поддержания синхронизации частот. Увеличение силы притягивающей связи делает такие периоды все более редкими, но не невозможными. Ниже из результатов данного раздела можно явно видеть, что совпадение средних частот при притягивающей связи и общем шуме не бывает идеальным.

Рассмотрим динамику отдельных осцилляторов в ансамбле. Удобно отслеживать сдвиг фазы $\ph_\varOmega-\varPhi$ отдельного осциллятора относительно фазы синхронизованного кластера. Дополним уравнения~(\ref{eq305}) и (\ref{eq306}) уравнением для сдвига фазы каждого отдельного осциллятора $\theta_\omega=\ph_\varOmega-\varPhi$ (см.\ уравнение~(\ref{eq104}));
\begin{align}
\dot{J}&=\mu_\beta J-2\gamma J(1+J)-\sigma\xi(t)\sqrt{J(1+J)}\cos\varPhi,
\label{eq403}\\[10pt]
\dot\varPhi&=\varOmega_0
-\mu\sin\beta\frac{J+1/2}{J+1}
+\sigma\xi(t)\frac{J+1/2}{\sqrt{J(1+J)}}\sin\varPhi,
\label{eq404}\\[10pt]
\dot\theta_\omega&=\omega -\mu\left(\sqrt{\frac{J}{1+J}}\sin(\theta_\omega+\beta) -\frac{J+1/2}{J+1}\sin\beta\right)+
\nonumber\\[5pt]
&\qquad\qquad+\sigma\xi(t)\left[\sin(\varPhi+\theta_\omega)
-\frac{J+1/2}{\sqrt{J(1+J)}}\sin\varPhi\right],
\label{eq405}
\end{align}
где $\omega=\varOmega-\varOmega_0$. В дальнейшем, для краткости записи, нижний индекс $\omega$ опускается.

Уравнение Фоккера--Планка для плотности вероятности состояний системы (\ref{eq403})--(\ref{eq405}) $W(J,\varPhi,\theta,t)$ имеет вид
\begin{gather}
\frac{\partial}{\partial t}W
+\frac{\partial}{\partial J}\bigg[\left(\mu_\beta J-2\gamma J(1+J)\right)W\bigg]
+\frac{\partial}{\partial\varPhi}\left[\left(\varOmega_0-\mu\sin\beta\frac{J+1/2}{J+1}\right)W\right] +{}
\nonumber
\\[5pt]
{}+\frac{\partial}{\partial\theta}\bigg[\bigg(\omega
-\mu\sqrt{\frac{J}{1+J}}\sin(\theta+\beta)
+\mu\frac{J+1/2}{J+1}\sin\beta
\bigg)W\bigg]
-\sigma^2\hat{Q}^2W=0\,,
\label{eq406}
\end{gather}
где оператор $\hat{Q}$ определен как
\begin{gather}
\hat{Q}(\cdot)\equiv\frac{\partial}{\partial J}
\bigg(-\sqrt{J(1+J)}\cos\varPhi\,(\cdot)\bigg)
+\frac{\partial}{\partial\varPhi}
\bigg(\frac{J+1/2}{\sqrt{J(1+J)}}\sin\varPhi\,(\cdot)\bigg) +{}
\nonumber
\\[5pt]
{}+\frac{\partial}{\partial\theta}
\bigg(\bigg(\sin(\varPhi+\theta)-\frac{J+1/2}{\sqrt{J(1+J)}}\sin\varPhi\bigg)(\cdot)\bigg)\,.
\label{eq407}
\end{gather}

\subsection{Распределение сдвига фаз для высокочастотных колебаний}
При исчезающе малых $\mu$, $\sigma$ и $\gamma$, плотность распределения вероятности
$W(J,\varPhi,\theta,t)=(2\pi)^{-1}w(J,\theta,t)$, где
\[
\int_0^{+\infty}\d{J}\int_0^{2\pi}\d\theta\,w(J,\theta)=1\;.
\]
Тогда, при $\mu\sim\sigma^2\sim\gamma\ll\varOmega_0$, можно считать $\mu=\sigma^2\mu_1$, $\gamma=\sigma^2\gamma_1$, $\omega=\sigma^2\omega_1$ и применить метод многих масштабов; $W=W^{(0)}(J,\theta,t_1,t_2,...)+\sigma^2W^{(1)}(J,\varPhi,\theta,t_0,t_1,t_2...)+...$\,, где $t_n=\sigma^{2n}t$. В ведущем порядке уравнение~(\ref{eq406}) дает $W^{(0)}=(2\pi)^{-1}w(J,\theta,t_1,t_2,...)$. В порядке $\sigma^2$ уравнение~(\ref{eq406}) дает
\[
\begin{gathered}
\frac{\partial W^{(1)}}{\partial t_0}+\varOmega_0\frac{\partial W^{(1)}}{\partial\varPhi}
+\frac{\partial W^{(0)}}{\partial t_1}
+\frac{\partial}{\partial J}\bigg[\big(\mu_{\beta,1}J-2\gamma_1J(1+J)\big)W^{(0)}\bigg]+
\\[5pt]
{}+\frac{\partial}{\partial\varPhi}\bigg[\bigg(-\mu_1\sin\beta\frac{J+1/2}{J+1}\bigg)W^{(0)}\bigg]+
\\[5pt]
{}+\frac{\partial}{\partial\theta}\bigg[\bigg(\omega_1
-\mu_1\sqrt{\frac{J}{1+J}}\sin(\theta+\beta)
+\mu_1\frac{J+1/2}{J+1}\sin\beta
\bigg)W^{(0)}\bigg]
-\hat{Q}^2W^{(0)}=0\,.
\end{gathered}
\]
Проинтегрировав полученное уравнение по $\varPhi$ от $0$ до $2\pi$, можно получить
\begin{equation}
\begin{gathered}
\frac{\partial}{\partial t_0}\int_0^{2\pi}W^{(1)}\d\varPhi
+\frac{\partial w(J,\theta,t_1)}{\partial t_1}
+\frac{\partial}{\partial J}\bigg[\big(\mu_{\beta,1}J-2\gamma_1J(1+J)\big)w(J,\theta,t_1)\bigg] +{}
\nonumber\\[5pt]
{}+\frac{\partial}{\partial\theta}\bigg[\bigg(\omega_1
-\mu_1\sqrt{\frac{J}{1+J}}\sin(\theta+\beta)
+\mu_1\frac{J+1/2}{J+1}\sin\beta\bigg)w(J,\theta,t_1)\bigg] -{}
\nonumber\\[5pt]
{}-\frac{1}{2\pi}\int_0^{2\pi}\hat{Q}^2w(R,\theta,t_1)\,\d\varPhi=0\,.
\end{gathered}
\end{equation}
Чтобы избежать линейного роста $W^{(1)}$ в ``нормальном'' времени $t_0$, что нарушило бы иерархию малости вкладов в разложении, необходимо потребовать обращение первого слагаемого в ноль. Полученный интеграл может быть записан в виде
\[
\frac{1}{2\pi}\int\limits_0^{2\pi}\d\varPhi\,\hat{Q}^2w(J,t_1)
=\frac{\partial}{\partial J}\left(-\frac{J+1/2}{2}w\right)
+\frac{\partial}{\partial\theta}
\left(\frac{\sin{\theta}}{2}\frac{J+1/2}{\sqrt{J(1+J)}}w\right)
+\hat{Q}_{J,\theta}^2w+\hat{Q}_{\theta}^2w\,,
\]
где
\[
\hat{Q}_{J,\theta}(\cdot)\equiv
\frac{\partial}{\partial J}\bigg(-\sqrt{\frac{J(1+J)}{2}}\,(\cdot)\bigg)
+\frac{\partial}{\partial\theta}\left(\frac{\sin{\theta}}{\sqrt{2}}(\cdot)\right),\quad
\hat{Q}_\theta\equiv
\frac{\partial}{\partial\theta}\left(\left(\frac{\cos{\theta}}{\sqrt{2}}
-\frac{J+1/2}{\sqrt{2J(1+J)}}\right)(\cdot)\right)\,.
\]
Тогда плотность распределения вероятности $w(J,\theta,t)$ определяется уравнением
\begin{equation}
\begin{gathered}
\frac{\partial w}{\partial t}
+\frac{\partial}{\partial J}\bigg(\bigg[\mu_\beta J-2\gamma J(1+J)
+\frac{\sigma^2}{2}(J+1/2)\bigg]w\bigg)
+\frac{\partial}{\partial\theta}\bigg[\bigg(\omega
-\mu\sqrt{\frac{J}{1+J}}\sin(\theta+\beta) +{}
\\[7pt]
{}+\mu\frac{J+1/2}{J+1}\sin\beta
-\frac{\sigma^2}{2}\frac{J+1/2}{\sqrt{J(1+J)}}\sin\theta
\bigg)w\bigg]
-\sigma^2\hat{Q}_{J,\theta}^2w-\sigma^2\hat{Q}_\theta^2w=0\,.
\end{gathered}
\label{eq4a01}
\end{equation}

Уравнение~(\ref{eq4a01}) может быть рассмотрено как уравнение Фоккера--Планка  для стохастической системы уравнений с двумя независимыми шумами  $\zeta_1(t)$ и $\zeta_2(t)$\,:
\begin{align}
\dot{J}&=\mu_\beta J-2\gamma J(1+J)+\frac{\sigma^2}{2}(J+1/2)
-\sigma\sqrt{\frac{J(1+J)}{2}}\zeta_1(t)\,,\quad
\label{eq4a02}
\\[10pt]
\dot{\theta}&=\omega
-\mu\sqrt{\frac{J}{1+J}}\sin(\theta+\beta)
+\mu\frac{J+1/2}{J+1}\sin\beta
-\frac{\sigma^2}{2}\frac{J+1/2}{\sqrt{J(1+J)}}\sin\theta +{}
\nonumber\\[5pt]
&\qquad\qquad
{}+\sigma\frac{\sin\theta}{\sqrt{2}}\zeta_1(t)
+\sigma\left(\frac{\cos{\theta}}{\sqrt{2}}
-\frac{J+1/2}{\sqrt{2J(1+J)}}\right)\zeta_2(t)\,.
\label{eq4a03}
\end{align}
Исходный шум $\xi(t)$ генерирует два независимых шума $\zeta_1(t)$ м $\zeta_2(t)$, которые являются гауссовскими и дельта-коррелированными, $\la\zeta_n(t)\zeta_l(t+t')\ra=2\delta_{n,l}\delta(t')$\,, так как сигналы $\xi(t)\cos{\varOmega_0t}$ и $\xi(t)\sin{\varOmega_0t}$ являются некоррелированными  на масштабе времени $2\pi/\varOmega_0$.

\subsection{Динамика состояний, близких к синхронным, при малой\\ расстройке частот}
Рассмотрим случай малой расстройки частот $\gamma\ll\sigma^2\sim|\mu|$ и $2\mu/\sigma^2>-1$. (Заметим, что по сравнению с $\varOmega_0$, $\gamma$ предполагается того же порядка малости, что и $\sigma^2$ и $\mu$, то есть, $\sigma^3\ll\gamma\ll\sigma^2$). В этом случае состояния системы собираются к $J\gg1$, что может быть использовано для интегрирования~(\ref{eq4a01}) по $J$. Для $\overline{w}\equiv\int_0^{+\infty}w\,\d{J}$, с $\overline{f(J)w}\approx\overline{w}\lim_{J\to+\infty}f(J)$, из уравнения~(\ref{eq4a01}) можно найти
\[
\begin{gathered}
\frac{\partial\overline{w}}{\partial t}
+\frac{\partial}{\partial\theta}\bigg(\left[\omega-\mu\big(\sin(\theta+\beta)-\sin{\beta}\big) -\frac{\sigma^2}{2}\sin{\theta}\right]\overline{w}\bigg) -{}
\\[5pt]
{}-\frac{\sigma^2}{2}\frac{\partial}{\partial\theta}\Big(\sin{\theta}
\frac{\partial}{\partial\theta}\Big(\sin{\theta}\,\overline{w}\Big)\Big)
-\frac{\sigma^2}{2}\frac{\partial}{\partial\theta}\Big((\cos{\theta}-1)
\frac{\partial}{\partial\theta}\Big((\cos{\theta}-1)\,\overline{w}\Big)\Big)=0\,.
\end{gathered}
\]
Упрощая, можно получить
\begin{equation}
\frac{\partial\overline{w}}{\partial t}
+\frac{\partial}{\partial\theta}\Big(\left[\omega
-\mu\big(\sin(\theta+\beta)-\sin{\beta}\big)\right]\overline{w}\Big)
-\sigma^2\frac{\partial^2}{\partial\theta^2}\Big((1-\cos\theta)\overline{w}\Big)=0\,.
\label{eq4b01}
\end{equation}

\begin{figure*}[!t]
	\centerline{
		{\sf (a)}\hspace{-5mm}
		\includegraphics[width=0.42\textwidth]%
		{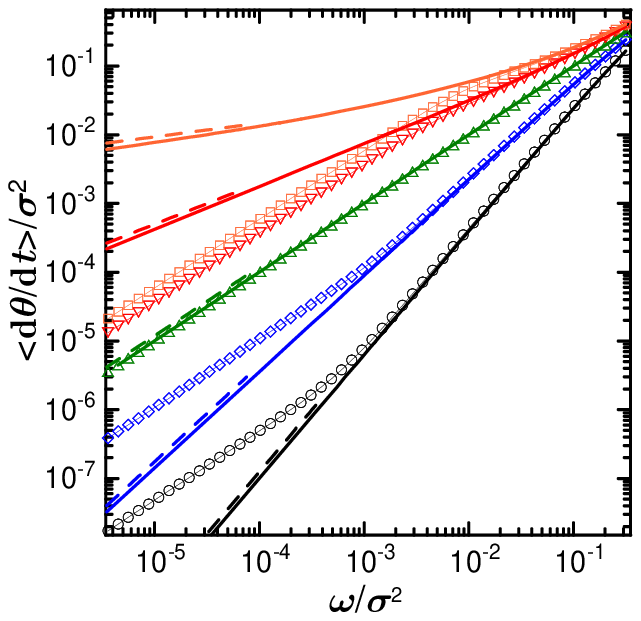}
		\qquad\quad
		{\sf (b)}\hspace{-5mm}
		\includegraphics[width=0.42\textwidth]%
		{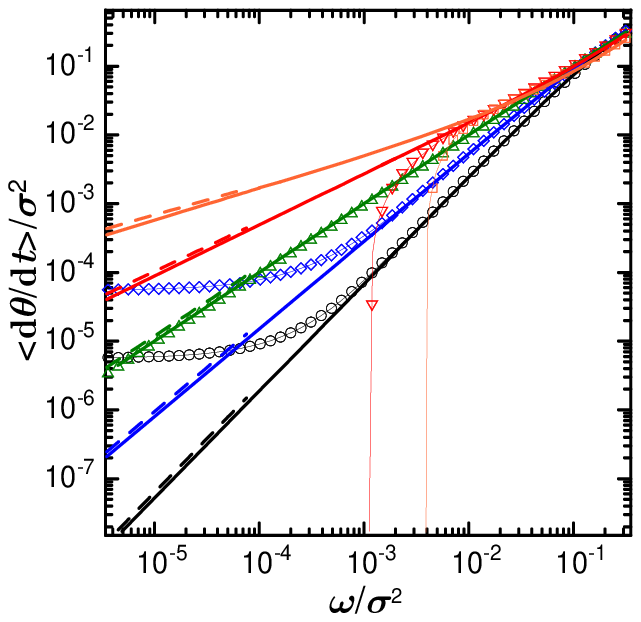}
	}
	\vspace{10pt}
	\centerline{
		{\sf (c)}\hspace{-5mm}
		\includegraphics[width=0.42\textwidth]%
		{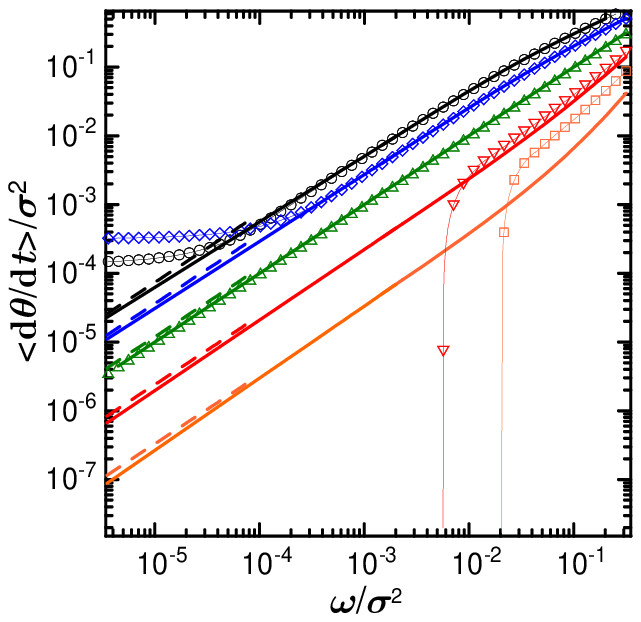}
		\qquad\quad
		{\sf (d)}\hspace{-5mm}
		\includegraphics[width=0.42\textwidth]%
		{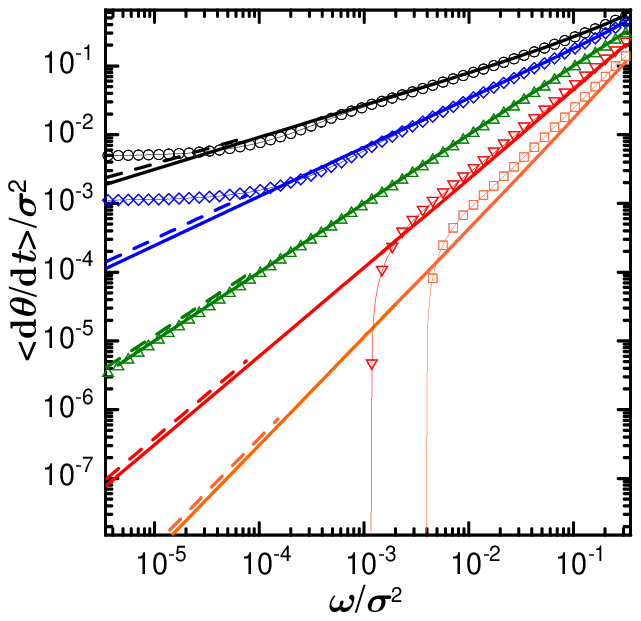}
	}
	
	\caption{
Зависимость средней частоты $\la\dot\theta\ra$ от расстройки частоты $\omega$ при $\beta=0$ (a), $\pi/4$ (b), $\pi/2$ (c), и $3\pi/4$ (d). Результаты численного счета с помощью цепных дробей при $\gamma=0.01$ построены на (a), (b), (d) окружностями, ромбами, треугольниками с вершинами вверх и вниз, квадратами ($\mu/\sigma^2=0.4$, $0.2$, $0$, $-0.2$, $-0.4$ соответственно). Результаты на графике (c) --- для $\mu/\sigma^2=0.8$, $0.4$, $0$, $-0.4$, $-0.8$. Сплошными линиями представлены результаты для $\gamma=0$ и указанных $\mu/\sigma^2$; пунктирными линиями показан наклон $1+2\mu\cos\beta/\sigma^2$, соответствующий асимптотическому поведению $\la\dot\theta\ra\sim\omega^{1+2\mu\cos\beta/\sigma^2}$. На графиках (b)--(d) некоторые кривые пересекают ноль не при $\omega=0$ --- в log--log масштабе это можно наблюдать в виде выходов на горизонтальное плато (когда $\la\dot\theta\ra_{\omega=0}>0$) или уходов кривых вниз (когда $\la\dot\theta\ra_{\omega=0}<0$); на рис.~\ref{fig4} зависимости представлены со смещением так, чтобы пересечение кривых с нулем происходило в начале координат.
	}
	\label{fig3}
\end{figure*}

\begin{figure*}[!t]
	\centerline{
		{\sf (a)}\hspace{-5mm}
		\includegraphics[width=0.42\textwidth]%
		{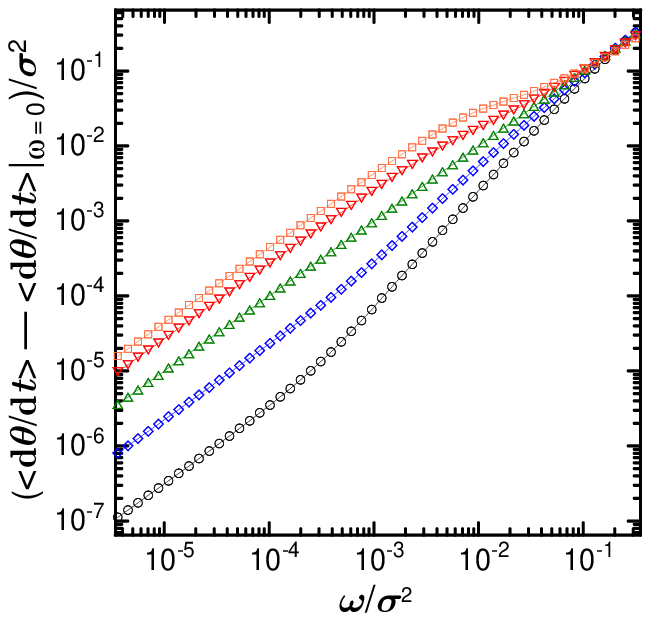}
		\qquad\quad
		{\sf (b)}\hspace{-5mm}
		\includegraphics[width=0.42\textwidth]%
		{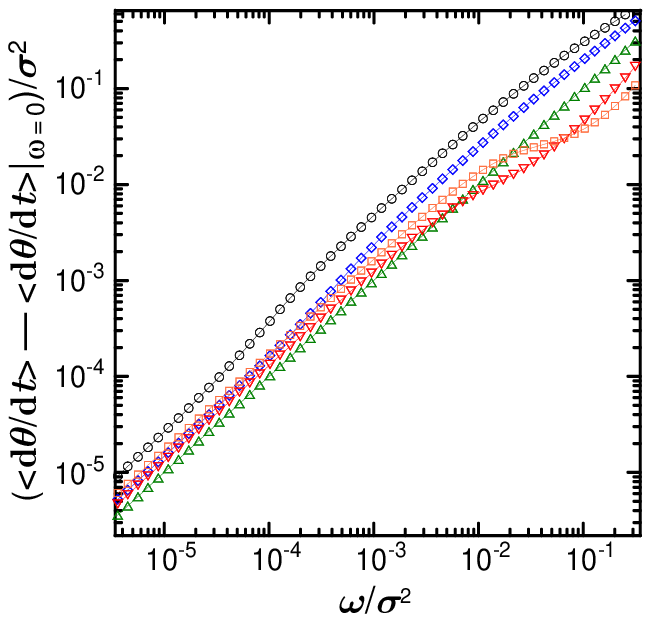}
	}
	\vspace{10pt}
	\centerline{
		{\sf (c)}\hspace{-5mm}
		\includegraphics[width=0.42\textwidth]%
		{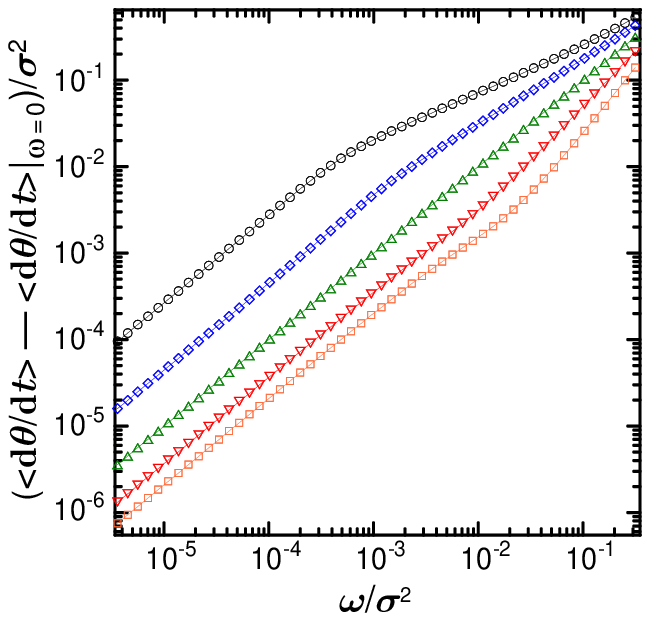}
	}
	
	\caption{
		Зависимость смещенной средней частоты $\la\dot\theta\ra-\la\dot\theta\ra_{\omega=0}$ от расстройки частоты $\omega$ при $\gamma=0.01$ и $\beta=\pi/4$ (a), $\pi/2$ (b), $3\pi/4$ (c); обозначения таки же, как и на рис.~\ref{fig3}
	}
	\label{fig4}
\end{figure*}

Рассмотрим статистически стационарное состояние, соответствующее постоянному по $\theta$ потоку вероятности $j$ в уравнении~(\ref{eq4b01}):
\begin{equation}
\Big(\omega
-\mu\big(\sin(\theta\!+\!\beta)-\sin{\beta}\big)\Big)\overline{w}
-\sigma^2\frac{\mathrm{d}}{\mathrm{d}\theta}\Big((1-\cos\theta)\overline{w}\Big)=j\;.
\label{eq4b02}
\end{equation}

В терминах последнего уравнения, захват частоты $\la\dot\theta\ra=0$ соответствует $j=0$. При $\sigma\ne0$ можно показать, что решение с $j=0$ существует для одного выделенного значения $\omega$, а это означает, что полный захват частоты невозможен даже при малых отклонениях собственной частоты от средней частоты ансамбля.

При $j\ne0$ находим
\begin{equation}
\overline{w}(\theta)
=\frac{j}{\sigma^2}
\int\limits_{\theta}^{2\pi}\d\psi\frac{(1-\cos\psi)^{\mu_\beta/\sigma^2}}{(1-\cos\theta)^{1+\mu_\beta/\sigma^2}}
\exp\left[-\frac{\omega}{\sigma^2}\left(\ctg\frac{\theta}{2}-\ctg\frac{\psi}{2}\right)
-\frac{\mu\sin\beta}{\sigma^2}(\psi-\theta)\right]\!.\;\;
\label{eq4b03}
\end{equation}
Это выражение имеет хорошие свойства сходимости при  $\theta=0$ и $2\pi$: $\overline{w}(0)=\overline{w}(2\pi)=j/\omega$; при ненулевом $\omega$ оно сходится при любых значениях $\mu$. Условие нормировки позволяет найти поток $j$ как функцию $\mu$, $\beta$, $\sigma^2$ и $\omega$:
\begin{eqnarray}
j=\sigma^2\,f\left(\frac{\mu}{\sigma^2},\frac{\omega}{\sigma^2},\beta\right)\,.
\label{eq4b04}
\end{eqnarray}

\subsection{Средняя частота осцилляторов в состояниях, близких\\ к синхронным}
Найдем среднюю частоту осцилляторов для случая неполной синхронизации. С технической точки зрения, задача сводится к вычислению средней частоты для уравнения~(\ref{eq4a03}) при конечном, но все еще большом $J$. Для приблизительных вычислений перепишем уравнение~(\ref{eq4a03}) в виде
\begin{align}
\dot\theta=\omega_\beta
-\mu b\sin(\theta+\beta)-\frac{\sigma^2}{2}c\sin\theta
-\frac{\sigma}{\sqrt{2}}\sin\theta\,\zeta_1(t)
+\frac{\sigma}{\sqrt{2}}(\cos\theta-c)\,\zeta_2(t)\,.
\label{eq4c01}
\end{align}
Рассмотрим полученное уравнение для постоянных коэффициентов $\omega_\beta$, $b$ и $c$, вычисленных как средние значения:
\begin{gather}
\omega_\beta\equiv\omega+\mu\lla\frac{J+1/2}{J+1}\rra\sin\beta
=\omega+\mu\left(1-\frac12\lla\frac{1}{J+1}\rra\right)\sin\beta\,,
\label{eq4c011}
\\[10pt]
b\equiv\lla\sqrt{\frac{J}{1+J}}\rra
=1-\frac12\lla\frac{1}{1+J}\rra
-\frac18\lla\frac{1}{(1+J)^2}\rra
+\dots\,,
\label{eq4c012}
\\[10pt]
c\equiv\lla\frac{J+1/2}{\sqrt{J(1+J)}}\rra
=1+\frac18\lla\frac{1}{(1+J)^2}\rra
+\dots\,.
\label{eq4c013}
\end{gather}
С учетом распределения~(\ref{eq3c03}), имеем
\begin{equation}
\lla\frac{1}{(1+J)^n}\rra=\left(\frac{4\gamma}{\sigma^2}\right)^n
\frac{\displaystyle\Gamma\left(\frac{2\mu_\beta}{\sigma^2}+1-n,\frac{4\gamma}{\sigma^2}\right)}
{\displaystyle\Gamma\left(\frac{2\mu_\beta}{\sigma^2}+1,\frac{4\gamma}{\sigma^2}\right)}\,.
\label{eq4c02}
\end{equation}

Уравнение~(\ref{eq4c01}) с постоянными коэффициентами дает уравнение Фок\-кера-Планка
\begin{gather}
\frac{\partial w(\theta,t)}{\partial t}
+\frac{\partial}{\partial\theta}\!\left[\left(\omega_\beta
-\mu b\sin(\theta\!+\!\beta)
-\frac{\sigma^2}{2}c\sin\theta\right)w(\theta,t)\right] -{}
\\[5pt]
 {}-\frac{\sigma^2}{2}\frac{\partial}{\partial\theta}\bigg(\sin\theta
\frac{\partial}{\partial\theta}\bigg(\sin\theta\,w(\theta,t)\bigg)\bigg)
-\frac{\sigma^2}{2}\frac{\partial}{\partial\theta}\bigg((\cos\theta-c)
\frac{\partial}{\partial\theta}\bigg((\cos\theta-c)\,w(\theta,t)\bigg)\bigg)=0\,,
\end{gather}
которое может быть приведено к упрощенному виду
\begin{align}
\frac{\partial w(\theta,t)}{\partial t}
+\frac{\partial}{\partial\theta}\bigg[\Big(\omega_\beta
-\mu b\sin(\theta+\beta)\Big)w(\theta,t)\bigg]
-\frac{\sigma^2}{2}\frac{\partial^2}{\partial\theta^2}
\bigg[\Big(1+c^2-2c\cos\theta\Big)\,w(\theta,t)\bigg]=0\,.
\label{eq4c03}
\end{align}
Для постоянного по времени распределения $w(\theta)$ последнее уравнение может быть проинтегрировано по $\theta$:
\begin{align}
(\omega_\beta-\mu b\sin(\theta+\beta))w
-\frac{\sigma^2}{2}\frac{\mathrm{d}}{\mathrm{d}\theta}\big[(1+c^2-2c\cos\theta)w\big]=j\,,
\label{eq4c04}
\end{align}
где $j=const$ --- константа интегрирования, соответствующая потоку вероятности в системе, $\la\dot\theta\ra=2\pi j$.

\subsubsection{Представление плотности вероятности в виде цепных дробей}
Для вычисления частот осцилляторов, воспользуемся Фурье-разложением для $w(\theta)$,
\[
w(\theta)=\frac{1}{2\pi}\sum_{k=-\infty}^{+\infty} w_k e^{ik\theta}\,,
\]
и подставим его в уравнение~(\ref{eq4c04}):
\[
\begin{gathered}
\textstyle
\Big(\omega_\beta-\mu b\frac{e^{i(\theta+\beta)}
	-e^{-i(\theta+\beta)}}{2i}\Big)\sum\limits_k w_k e^{ik\theta}
-\sigma^2\sum\limits_k ik w_k \frac{1+c^2}{2}e^{ik\theta}
+\sigma^2c\frac{d}{d\theta}\sum\limits_k\frac{e^{i\theta}+e^{-i\theta}}{2}w_k e^{ik\theta}=2\pi j\,.
\end{gathered}
\]
Перегруппируем слагаемые:
\[
\begin{gathered}
\textstyle
\sum\limits_k e^{ik\theta}\bigg[\omega_\beta w_k
-\frac{\mu b\,e^{i\beta}}{2i}w_{k-1}
+\frac{\mu b\,e^{-i\beta}}{2i}w_{k+1}
-ik\sigma^2\frac{1+c^2}{2} w_k
+\frac{ik\sigma^2 c}{2}(w_{k-1}+w_{k+1})\bigg]=2\pi j\,.
\end{gathered}
\]
Для удобства рассмотрим отдельно $k=0$ и $k>0$, при этом будем иметь в виду, что $w_0=1$, $w_{-k}=w_k^*$. Для $k=0$,
\[
\omega_\beta-\frac{\mu b\,e^{i\beta}}{2i}w_1^*
+\frac{\mu b\,e^{-i\beta}}{2i}w_1=2\pi j
\]
или
\begin{equation}
2\pi j=\la\dot{\theta}\ra=\omega_\beta
+\mu b\,\mathrm{Im}(w_1e^{-i\beta})\,,
\label{eq4c05}
\end{equation}
для $k\ge1$,
\begin{align}
w_{k-1}\left[-\frac{\mu b\,e^{i\beta}}{2i}
+\frac{ik\sigma^2 c}{2}\right]
+w_k\left[\omega_\beta-ik\sigma^2\frac{1+c^2}{2}\right]
+w_{k+1}\left[\frac{\mu b\,e^{-i\beta}}{2i}+\frac{ik\sigma^2 c}{2}\right]=0\,.
\label{eq4c06}
\end{align}

Введем отношение $w_{k-1}/w_k=r_k$. Оно может быть выражено из уравнения~(\ref{eq4c06}):
\begin{equation}
r_k=\frac{\displaystyle 1+c^2+\frac{2i\omega_\beta}{k\sigma^2}}
{\displaystyle c+\frac{\mu b\,e^{i\beta}}{k\sigma^2}}
-\frac{\displaystyle c-\frac{\mu b\,e^{-i\beta}}{k\sigma^2}}
{\displaystyle c+\frac{\mu b\,e^{i\beta}}{k\sigma^2}}
\frac{1}{r_{k+1}}\,.
\label{eq4c07}
\end{equation}

Уравнение~(\ref{eq4c07}) представляет собой рекуррентное соотношение, позволяющее вычислить $w_1$ в виде цепной дроби
\begin{equation}
w_1=\frac{1}{\displaystyle A_1
	-B_1\frac{1}{\displaystyle A_2
		-B_2\frac{1}{\displaystyle A_3
			-B_3\frac{1}{\dots}}}}\,,
\label{eq4c08}
\end{equation}
где
\[
A_k\equiv\frac{\displaystyle 1+c^2+\frac{2i\omega_\beta}{k\sigma^2}}
{\displaystyle c+\frac{\mu b\,e^{i\beta}}{k\sigma^2}}
\quad
\mbox{   и   }
\quad
B_k\equiv\frac{\displaystyle c-\frac{\mu b\,e^{-i\beta}}{k\sigma^2}}
{\displaystyle c+\frac{\mu b\,e^{i\beta}}{k\sigma^2}}\,.
\]
Для достижения лучшей сходимости оборванного ряда цепных дробей, можно заметить, что $A_\infty=1/c+c$, $B_\infty=1$ и, в соответствии с уравнением~(\ref{eq4c07}), $r_\infty=c$ (второе решение $r_\infty=1/c$ соответствует расходящемуся разложению при $c>1$). Ошибка аппроксимации минимальна при $r_\infty$, подставленным на место $r_k$ в уравнение~(\ref{eq4c08}). Найдя $w_1$, можно вычислить среднюю частоту из уравнения~(\ref{eq4c05}). Результаты расчета $\la\dot{\theta}\ra$ представлены на рис.~\ref{fig3} и \ref{fig4}. Случай $\beta\in[\pi,2\pi)$ не требует дополнительного рассмотрения в силу симметрии $(\mu,\beta)\leftrightarrow(-\mu,\beta+\pi)$.

\subsubsection{Асимптотическое поведение средних частот осцилляторов}
Рассмотрим взаимное притяжение и отталкивание средних частот при малых $\omega$ в уравнении~(\ref{eq4c04}). Перепишем уравнение~(\ref{eq4c04}) в терминах $G(\theta)\equiv(1+c^2-2c\cos\theta)\,w(\theta)$:
\begin{equation}
\frac{\omega_\beta-\mu b\sin(\theta+\beta)}{1+c^2-2c\cos\theta}G
-\frac{\sigma^2}{2}\frac{\mathrm{d}}{\mathrm{d}\theta}G=j\,.
\label{eq4c09}
\end{equation}
Для начала, найдем собственную частоту $\omega_0$ (или $\omega_{\beta,0}$, которая связана с $\omega_0$ уравнением~(\ref{eq4c011})) при $j=0$. Для этой частоты уравнение~(\ref{eq4c09}) приобретает вид
\begin{equation}
\frac{\omega_{\beta,0}-\mu b\sin(\theta+\beta)}{1+c^2-2c\cos\theta}G_0
-\frac{\sigma^2}{2}\frac{\mathrm{d}}{\mathrm{d}\theta}G_0=0\,.
\label{eq4c10}
\end{equation}
Разделив последнее уравнение на $G_0$ и проинтегрировав, можно получить
\begin{gather}
const-\frac{\mu b\cos\beta}{2c}\ln(1+c^2-2c\cos\theta)
+\frac{\mu b\sin\beta}{2c}\theta +{}
\nonumber
\\[5pt]
{}+\left(\omega_{\beta,0}-\mu b\sin\beta\frac{1+c^2}{2c}\right)\frac{2}{c^2-1}\arctg\left(\frac{c+1}{c-1}\tg\frac{\theta}{2}\right)
-\frac{\sigma^2}{2}\ln{G_0(\theta)}=0\,.
\label{eq4c11}
\end{gather}
Условие периодичности для $G_0(\theta)$ требует ${G_0(\pi-0)}={G_0(-\pi+0)}$, и уравнение~(\ref{eq4c11}) дает
$
\omega_{\beta,0}=\mu\frac{b}{c}\sin\beta
$.
Подставляя выражения (\ref{eq4c011})--(\ref{eq4c013}), находим
\begin{equation}
\omega_0=-\frac{\mu}{4}\left\langle(1+J)^{-2}\right\rangle
\Big(1+\mathcal{O}\left(\left\langle(1+J)^{-1}\right\rangle\right)\Big)\sin\beta\,.
\label{eq4c13}
\end{equation}
Следует заметить, что для состояния, близкого к синхронизации, при котором $J$ не стремится к бесконечности, собственная частота осциллятора не равна нулю при $\sin\beta\ne0$. Ее отклонение от нуля тем сильнее, чем слабее синхронизация. Без учета нормировки, периодическое решение  $G_0(\theta)$ имеет вид
\begin{align}
G_0(\theta)=
(1+c^2-2c\cos\theta)^{-\frac{\mu b\cos\beta}{\sigma^2c}}
\exp\left[\frac{2\mu b\sin\beta}{\sigma^2c}
\left\{\frac{\theta}{2}
-\arctg\left(\frac{c+1}{c-1}\tg\frac{\theta}{2}\right)
\right\}\right].
\label{eq4c14}
\end{align}
После нормировки решение принимает вид $\left\|w_0\right\|^{-1}G_0(\theta)$, где
$
\left\|...\right\|\equiv\int_{-\pi}^{\pi}...\,\mathrm{d}\theta\,,
$
и $w_0(\theta)=G_0(\theta)/(1+c^2-2c\cos\theta)$\,.

Вычислим $j=\epsilon j_1$ для $\omega_\beta=\omega_{\beta,0}+\epsilon\omega_{\beta,1}$, где $\epsilon\ll1$. Используя разложение $G=\left\|w_0\right\|^{-1}G_0+\epsilon G_1+...$, находим в ведущем порядке ($\epsilon^1$) уравнения~(\ref{eq4c09})
\begin{equation}
\frac{\displaystyle
	\omega_{\beta,1}\frac{G_0}{\left\|w_0\right\|}+\big(\omega_{\beta,0}-\mu b\sin(\theta+\beta)\big)G_1}{1+c^2-2c\cos\theta}
-\frac{\sigma^2}{2}\frac{\mathrm{d}G_1}{\mathrm{d}\theta}=j_1\,.
\nonumber
\end{equation}
Умножая последнее уравнение на решение Эрмитово-сопряженной задачи $G_0^+(\theta)$ для уравнения~(\ref{eq4c10}) и интегрируя результат по $\theta$, находим
\begin{equation}
\frac{1}{\left\|w_0\right\|}\left\|\frac{G_0^+G_0}{1+c^2-2c\cos\theta}\right\|\omega_{\beta,1}
=j_1\left\|G_0^+\right\|\,.
\label{eq4c15}
\end{equation}
Эрмитово-сопряженная задача имеет вид
\begin{equation}
\frac{\omega_{\beta,0}-\mu b\sin(\theta+\beta)}{1+c^2-2c\cos\theta}G_0^+
+\frac{\sigma^2}{2}\frac{\mathrm{d}}{\mathrm{d}\theta}G_0^+=0\,,
\nonumber
\end{equation}
и, без учета нормировки,
\begin{align}
G_0^+(\theta)=
(1+c^2-2c\cos\theta)^{\frac{\mu b\cos\beta}{\sigma^2c}}
\exp\left[-\frac{2\mu b\sin\beta}{\sigma^2c}
\left\{\frac{\theta}{2}
-\arctg\left(\frac{c+1}{c-1}\tg\frac{\theta}{2}\right)
\right\}\right].
\label{eq4c16}
\end{align}
Тогда, с учетом  того, что $G_0(\theta)\,G_0^+(\theta)=1$ и $\left\|(1+c^2-2c\cos\theta)^{-1}\right\|=2\pi/(c^2-1)$, уравнение~(\ref{eq4c15}) принимает упрощенный вид:
\begin{equation}
j_1=\frac{2\pi}{(c^2-1)\left\|G_0^+\right\|\left\|w_0\right\|}\,\omega_{\beta,1}\,.
\label{eq4c17}
\end{equation}
Интегралы в последнем выражении могут быть вычислены следующим образом:
\[
\left\|G_0^+\right\|=\int\limits_{-\pi}^{\pi}(1+c^2-2c\cos\theta)^m e^{-2\alpha f(\theta)} \mathrm{d}\theta\,,\quad
\left\|w_0\right\|=\int\limits_{-\pi}^{\pi}(1+c^2-2c\cos\theta)^{-m-1} e^{2\alpha f(\theta)} \mathrm{d}\theta\,,
\]
где $m\equiv\mu b\cos\beta/(\sigma^2c)$, $\alpha\equiv \mu b\sin\beta/(\sigma^2c)$ и $f(\theta)\equiv\theta/2-\arctg\{[(c+1)/(c-1)]\tg(\theta/2)\}$.
При $(c-1)\ll1$ и $\mu\cos\beta/\sigma^2>-1/2$, что является условием устойчивости синхронного состояния, полученные интегралы можно упростить. В частности, для $m>-1/2$ и $(c-1)\ll1$, имеем
\[
\left\|G_0^+\right\|\approx
2\int\limits_0^\pi(1+c^2-2c\cos\theta)^m \ch\left[\alpha(\pi-\theta)\right]\mathrm{d}\theta
\]
(так как первый множитель делает вклад окрестности $\theta=0$ в интеграл малым, можно пренебречь отклонением $f(\theta)$ от $\mathrm{sign}(\theta)(|\theta|-\pi)/2$), которое не является малым только при $\theta\sim(c-1)$), и
\begin{align}
\left\|w_0\right\|\approx
2\int\limits_0^\pi[(c-1)^2+c\,\theta^2]^{-m-1} \ch\left[2\alpha f(\theta)\right]\mathrm{d}\theta
\approx
2\int\limits_0^\pi \frac{\ch\left[-2\alpha\arctg\left(\frac{\theta}{c-1}\right)\right]}
{[(c-1)^2+c\,\theta^2]^{m+1}}\mathrm{d}\theta
\nonumber
\end{align}
(в этом случае принципиальный вклад вносит как раз малая окрестность $\theta=0$, и можно использовать приближение, справедливое для $\theta\sim(c-1)$, в частности, $f(\theta)|_{\theta\sim(c-1)}\approx-\arctg[\theta/(c-1)]$).
Также,
\begin{align}
\left\|G_0^+\right\|\approx
2\int\limits_0^\pi2^m(1-\cos\theta)^m \ch\left[\alpha(\pi-\theta)\right]\mathrm{d}\theta
\approx
2^{2m+2}\int\limits_0^{\pi/2}\cos^{2m}\psi\, \ch(2\alpha\psi)\,\mathrm{d}\psi\,,
\label{eq4c18}
\end{align}
и, с учетом $\tg{y}=\theta/(c-1)$,
\begin{align}
\left\|w_0\right\|\approx
\frac{2}{(c-1)^{2m+1}}\int\limits_0^{\pi/2}
\frac{\ch(-2\alpha y)}{(1+\tg^2{y})^m}\,
\mathrm{d}y
\approx
\frac{2}{(c-1)^{2m+1}}\int\limits_0^{\pi/2}
\cos^{2m}\psi\,\ch(2\alpha\psi)\,
\mathrm{d}\psi\,.
\label{eq4c19}
\end{align}
Вычисляя интеграл, находим
\begin{align}
\int\limits_0^{\pi/2}
\cos^{2m}\psi\,\ch(2\alpha\psi)\,
\mathrm{d}\psi=
\frac{i\,\Gamma(2m+1)\,\Gamma(i\alpha-m)}{2^{2m+1}\,\Gamma(i\alpha+m+1)}
\sh[(\alpha+im)\pi]\,,
\nonumber
\end{align}
где $i$ --- мнимая единица. Подставляя выражения~(\ref{eq4c18}) и (\ref{eq4c19}) в уравнение~(\ref{eq4c17}), имеем
\begin{equation}
j_1=\frac{\pi}{2}
\left[\frac{i\,2^{m}(c-1)^{m}\Gamma(i\alpha+m+1)}
{\Gamma(2m+1)\,\Gamma(i\alpha-m)\sh[(\alpha+im)\pi]}
\right]^2
\omega_{\beta,1}\,.
\label{eq4c20}
\end{equation}
Окончательно, подставляя (\ref{eq4c013}), можно переписать уравнение~(\ref{eq4c20}) в виде
\begin{align}
\langle\dot{\theta}\rangle=
\left[\frac{\pi\,
	\Gamma\!\left(\frac{\mu e^{i\beta}}{\sigma^2}+1\right)}
{\Gamma\!\left(\frac{2\mu\cos\beta}{\sigma^2}+1\right)\,
	\Gamma\!\left(-\frac{\mu e^{-i\beta}}{\sigma^2}\right)\, \sin\!\left(\frac{\mu e^{-i\beta}}{\sigma^2}\pi\right)}
\right]^2
\left(\frac{\langle(1+J)^{-2}\rangle}{4}\right)^\frac{2\mu\cos\beta}{\sigma^2}
(\omega-\omega_0)\,,
\label{eq4c21}
\end{align}
где $\omega_0$ и среднее значение $\langle(1+J)^{-2}\rangle$ определяются уравнениями~(\ref{eq4c13}) и (\ref{eq4c02}) соответственно.
Заметим, что величина первого множителя в уравнении~(\ref{eq4c21}) имеет порядок $1$, тогда как $\langle(1+J)^{-2}\rangle$ мало при $\mu\cos\beta/\sigma^2>-1/2$ и, поэтому, коэффициент пропорциональности между $\langle\dot{\theta}\rangle$ и $(\omega-\omega_0)$ является малым для синхронизирующей связи ($\mu\cos\beta>0$) и большим для десинхронизирующей связи ($\mu\cos\beta<0$). Для предельного случая $\gamma\to0$, в котором $\langle(1+J)^{-2}\rangle\to0$, коэффициент стремится к нулю и бесконечности для синхронизирующей и десинхронизирующей связей соответственно.

\begin{figure*}[!t]
	\centerline{
		\includegraphics[width=0.60\textwidth]%
		{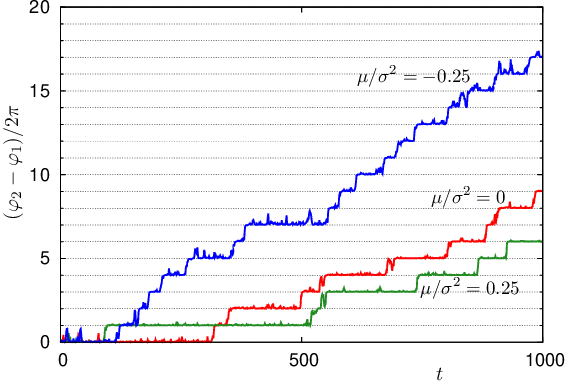}
	}
	
	\caption{Результаты численного счета для ансамбля $41$ осциллятора с гауссовским распределением собственных частот. На графике представлена динамика разности фаз двух осцилляторов в ансамбле при различной силе связи.
	}
	\label{fig5}
\end{figure*}

\paragraph{Физическая интерпретация для средних частот в ансамбле неидентичных осцилляторов.} Средние частоты всех осцилляторов с отличными собственными частотами различны. Для притягивающей связи средние частоты становятся ближе (притягиваются друг к другу). Для отталкивающей связи частоты начинают различаться сильнее (отталкиваются). Качественно, данный эффект можно объяснить следующим образом. В отсутствие связи частоты под действием шума остаются неизменными, хотя фазы стремятся сформировать кластер; в пределах кластера мгновенные частоты почти идентичны. Однако, присутствие отдельных осцилляторов в кластере является временным: фазы остаются вместе на протяжении некоторого времени, а затем фаза более быстрого осциллятора совершает дополнительный оборот относительно медленного осциллятора. Этот оборот в точности компенсирует близость мгновенных частот на протяжении периода совместной динамики осцилляторов в кластере. В случае притягивающей связи такие обороты становятся более редкими, так что они лишь частично компенсируют близость мгновенных частот на протяжении кластерных этапов динамики. Идеальной синхронизации частот соответствовало бы полное отсутствие оборотов, чего не наблюдается при наличии шума. В случае отталкивающей связи, дополнительные обороты становятся более частыми, так что происходит ``сверхкомпенсация'' близости мгновенных частот во время кластерных периодов. Данная картина проиллюстрирована на рис.~\ref{fig5}.

\section{Заключение}
В настоящей работе представлена теория синхронизации ансамбля фазовых осцилляторов под действием общего шума и глобальной связи. Взаимодействие глобальной связи и общего шума рассмотрено в рамках подхода Отта-Антонсена, позволяющего получить замкнутую систему стохастических уравнений для параметра порядка. В работе представлены аналитические решения полученных уравнений в тех случаях, когда это возможно. Отдельно рассмотрена ситуация, когда осцилляторы имеют достаточно большую собственную частоту, что позволяет произвести осреднение по быстро вращающейся фазе и получить более простые стохастические уравнения.

Наиболее интересные эффекты появляются тогда, когда общий шум пытается синхронизировать систему, а отталкивающая связь --- десинхронизировать. Для идентичных осцилляторов синхронизирующее воздействие шума преобладает над действием умеренной отталкивающей связи, так как полностью синхронное состояние является адсорбирующим, тогда как максимально асинхронное состояние является не адсорбирующим, а лишь слабо притягивающим. Для неидентичных осцилляторов состояние полной синхронизации невозможно --- вместо него возникает состояние неидельной синхронизации, которое перестает быть адсорбирующим, а становится слабо притягивающим, как и состояние максимальной асинхронности.

Для осцилляторов с расстройкой собственных частот наибольший интерес представляет поведение средних частот. Эффекты, связанные с общим шумом и глобальной связью, взаимодействуют нетривиальным образом. В отсутствии шума притягивающая связь собирает частоты вместе, в то время как отталкивающая связь не влияет на них. Шум в отсутствии связи также не оказывает влияния на частоты. Однако, когда на систему воздействует и шум, и отталкивающая связь, наблюдается расталкивание частот, при этом параметр порядка может оставаться достаточно большим.

\vspace{0.5cm}

{\em
Результаты, представленные в разделах 1.1-2, 2.1-2, 3.1-2, получены А.В.Д.\ и Д.С.Г.\ при финансовой поддержке Российского научно фонда (грант №~14-21-00090).
Результаты, представленные в разделах 1.3, 2.3 и 3.3, получены М.Р.\ и А.П.\ при финансовой поддержке Российского научного фонда (грант №~17-12-01534).
}

\section*{Приложение: Теории Отта-Антонсена (Ott-Antonsen) и Ватанабе-Строгаца (Watanabe-Strogatz)}
Широкий и важный класс систем, ставших парадигматическими для изучения коллективных эффектов в ансамблях фазовых осцилляторов, описывается уравнениями вида
\begin{equation}
\dot\varphi_k=\varOmega(t)+\mathrm{Im}(H(t)\,e^{-i\varphi_k}),
\qquad k=1,...,N\,,
\label{eqap01}
\end{equation}
где $\varOmega(t)$ и $H(t)$ --- произвольные функции времени. В самом деле, в термодинамическом пределе $N\gg1$ именно к таким системам относится ансамбль Курамото:
\[
\dot\varphi_k=\varOmega +\frac{\mu}{N}\sum_{j=1}^N\sin(\varphi_j-\varphi_k)\,,
\]
для которого можно полагать вид (\ref{eqap01}) с $H(t)=\mu N^{-1}\sum_{j=1}^Ne^{i\varphi_j}$, поскольку относительный индивидуальный вклад $k$-го осциллятора в $H(t)$ исчезающе мал. Для цепочки идентичных джозефсоновских (сверхпроводящих) контактов с шунтированием, выполняющим функцию глобальной связи, могут быть получены безразмерные уравнения~\cite{Marvel-Strogatz-2009}
\[
\dot\varphi_k=I_\mathrm{inp}(t)-I_0\sin\varphi_k+\frac{\mu}{N}{\textstyle\sum_{j=1}^N\sin\varphi_j}\,,
\]
где $I_\mathrm{inp}(t)$ --- ток, подаваемый в систему, $I_0$ --- параметр контактов, а фаза $\varphi_k$ по своему физическому смыслу является характеристикой волны сверхпроводящего тока и связана с легко наблюдаемой макроскопической величиной --- разностью потенциалов $U_k$ на контакте двух сверхпроводников: в размерном виде $\dot\varphi_k=(2e/\hbar)U_k$, $e$ --- элементарный заряд, $\hbar$ --- постоянная Планка. Такому ансамблю контактов соответствует
$\varOmega(t)=I_\mathrm{inp}(t)-\mu\,\mathrm{Im}(N^{-1}\sum_{j=1}^Ne^{i\varphi_j})$ и $H(t)=I_0$. Для системы связанных активных ротаторов (похожие уравнения применимы и для описания тета-нейронов)~\cite{Shinomoto-Kuramoto-1986} уравнения
\[
\dot\varphi_k=\varOmega+\frac{K}{N}\sum_{j=1}^N\sin(\varphi_j-\varphi_k)-B\sin\varphi_k\,,
\]
имеют вид~(\ref{eqap01}) с $H(t)=B+KN^{-1}\sum_{j=1}^Ne^{i\varphi_j}$.

Для уравнения (\ref{eqap01}) может быть записано Мастер-уравнение, определяющее эволюцию плотности вероятности $w(\varphi,t)$\,:
\begin{equation}
\frac{\partial w}{\partial t}
 +\frac{\partial}{\partial\varphi}\Big((\varOmega(t)-i\frac{H(t)}{2}e^{-i\varphi} +i\frac{H^\ast(t)}{2}e^{i\varphi})w\Big)
 =0\,.
\label{eqap02}
\end{equation}
В Фурье-пространстве, $w(\varphi,t)=(2\pi)^{-1}\big[1+\sum_{j=1}^{\infty}(a_je^{ij\varphi}+c.c.)\big]$ и Мастер-уравнение принимает вид:
\begin{equation}
\dot{a}_{j}=-ji\varOmega(t)\,a_j+j\frac{H^\ast(t)}{2}{a}_{j-1}-j\frac{H(t)}{2}{a}_{j+1}\,,
\label{eqap03}
\end{equation}
где $j=1,2,...$ и следует полагать $a_0=1$. Можно заметить, что последнее уравнение допускает частное решение вида $a_j=a_1^j$. В самом деле, при такой подстановке можно получить для $j=1$
\begin{equation}
\dot{a}_1=-i\varOmega(t)\,a_1+\frac{H^\ast(t)}{2}-\frac{H(t)}{2}a_1^2
\label{eqap04}
\end{equation}
и для $j>1$ ---
\[
ja_1^{j-1}\dot{a}_1=ja_1^{j-1}\left(-i\varOmega(t)\,a_1+\frac{H^\ast(t)}{2}-\frac{H(t)}{2}a_1^2\right)\,,
\]
что идентично уравнению (\ref{eqap04}). Таким образом, оказывается что динамика системы может происходить на некотором множестве, характеризующемся свойством $a_j=a_1^j$ и получившем название множества Отта-Антонсена (О-А)~\cite{Ott-Antonsen-2008}; динамика системы на этом множестве описывается обыкновенным дифференциальным уравнением для комплексной переменной $a_1$. Следует заметить связь между $a_1$ и комплексным параметром порядка $Z=\langle{e^{i\varphi}}\rangle=\int_0^{2\pi}e^{i\varphi}w(\varphi,t)\,\mathrm{d}\varphi=a_1^\ast(t)$. В итоге, на множестве О-А динамика системы исчерпывающе описывается ноль-мерным уравнением для параметра порядка $Z$:
\begin{equation}
\dot{Z}=i\varOmega(t)\,Z+\frac{H(t)}{2}-\frac{H^\ast(t)}{2}Z^2\,.
\label{eqap05}
\end{equation}
При этом для конкретных задач, как правило, имеет место простая функциональная связь между $\{\varOmega(t),H(t)\}$ и $Z$. В самом деле, применительно к трем упомянутым выше примерам: $H=\mu Z$ для ансамбля Курамото, $\varOmega=I_\mathrm{inp}(t)-\mu\,\mathrm{Im}(Z)$ и $H(t)=I_0$ для цепочки джозефсоновских контактов с шунтированием, $H=B+KZ$ для ансамбля активных ротаторов.

Важен вопрос об устойчивости множества О-А. Строго говоря, для идеально идентичных осцилляторов множество О-А является нейтрально устойчивым, а динамика системы, как было показано Ватанабе, Строгацом и др.~\cite{Watanabe-Strogatz-1994,Pikovsky-Rosenblum-2008,Marvel-Mirollo-Strogatz-2009}, полностью интегрируемой. Вместе с тем, в реальных системах имеет место неточное совпадение собственных частот осцилляторов, что приводит к появлению слабой устойчивости множества О-А и позволяет получать на основе уравнения (\ref{eqap05}) результаты, хорошо согласующиеся с результатами численного счета (см., например,~\cite{Braun-etal-2012}). Для объяснения механизма возникновения устойчивости множества О-А целесообразно коротко, но строго разобрать случай ансамбля идеально идентичных осцилляторов (Ватанабе-Строгац), для которого даже не требуется переход к термодинамическому пределу большого числа осцилляторов.

Для набора фаз $\varphi_k$ можно определить взаимно однозначное преобразование к набору фаз $\psi_k$, параметризуемое комплексным числом $z$:
\begin{equation}
e^{i\varphi_k}=\frac{z+e^{i\psi_k}}{1+z^\ast e^{i\psi_k}}
\quad\left(\mbox{и обратное: }\;
e^{i\psi_k}=\frac{e^{i\varphi_k}-z}{1-z^\ast e^{i\varphi_k}}\right)\,.
\label{eqap06}
\end{equation}
Если потребовать для новых фаз выполнение условия $\sum_{k=1}^{N}e^{i\psi_k}=0$, то преобразование становится единственным и однозначно определяет $z$.
Подстановка преобразования (\ref{eqap06}) в уравнение (\ref{eqap01}) позволяет получить уравнения эволюции $z$ и $\psi_k$, которые при условии $\sum_{k=1}^{N}e^{i\psi_k}=0$ принимают вид:
\begin{align}
\dot{z}&=i\varOmega(t)\,z+\frac{H(t)}{2}-\frac{H^\ast(t)}{2}z^2\,,
\label{eqap07}
\\
\dot{\psi}_k&=\varOmega(t)+\mathrm{Im}(H(t)\,z^\ast)\,.
\label{eqap08}
\end{align}
Можно заметить, что уравнение динамики (\ref{eqap07}) для переменной $z$, которая может быть введена в общем случае (не только на множестве О-А), эквивалентно уравнению для параметра порядка $Z$ на множестве О-А (\ref{eqap05}). (Процедура введения $z$ при этом такова, что из нее не следует в общем случае эквивалентность $z$ и $Z$.) Согласно уравнению (\ref{eqap08}), все фазы $\psi_k$ меняются одинаково. Таким образом, динамика системы определяется изменением $z$ и любой одной из фаз $\psi_{k^\prime}$, а разности всех фаз $(\psi_k-\psi_{k^\prime})$ являются константами, интегралами движения. Плотность вероятности $W(\psi_k,t)$ является бегущей волной с замороженным профилем.

Установим, чему соответствует решение О-А
\[
w_\mathrm{OA}(\varphi,t) =\frac{1}{2\pi}\Big[1+\sum_{j=1}^\infty([a_1(t)]^je^{ij\varphi}+c.c.)\Big] =\frac{1-|a_1|^2}{2\pi|1-a_1e^{i\varphi}|^2}=\frac{1-|Z|^2}{2\pi|1-Ze^{-i\varphi}|^2}
\]
в терминах $\{z,\psi_k\}$.
Поскольку для этого решения $Z$ подчиняется уравнению (\ref{eqap05}), и тому же уравнению всегда подчиняется переменная $z$, естественно предположить их совпадение в этом случае. Из соотношений (\ref{eqap06}) может быть вычислено
\[
\mathrm{d}\psi=\frac{1-|z|^2}{|1-ze^{-i\varphi}|^2}\mathrm{d}\varphi\,,
\]
откуда $W_\mathrm{OA}(\psi)=w_\mathrm{OA}(\varphi)\frac{|d\varphi|}{|d\psi|}=(2\pi)^{-1}$. Предположение о тождественности $z$ и $Z$ для решения О-А можно теперь проверить строго: для $W(\psi)=(2\pi)^{-1}$ и $z$ могут быть вычислены параметр порядка $Z$, который оказывается равен $z$, и $w(\varphi)$, оказывающееся совпадающим с $w_\mathrm{OA}(\varphi|Z=z)$. Таким образом получаем, что решение Отта-Антонсена --- это частный случай, соответствующий равномерному распределению фаз $\psi_k$.

Теперь, когда установлено соответствие между решением О-А и описанием системы в терминах $\{z,\psi_k\}$, можно рассмотреть вопрос о том, что при неидеальной идентичности осцилляторов множество О-А из нейтрально устойчивого превращается в ``слабо'' притягивающее. В самом деле, в ансамбле с расстройкой собственных частот можно выделить подгруппы осцилляторов с одинаковыми частотами. Распределение $W(\psi,t)$ для каждой такой подгруппы является бегущей волной с замороженным профилем. Однако, из-за разностей частот эти волны, согласно уравнению (\ref{eqap08}), бегут с различной скоростью, что приводит к их постоянному смешению. В термодинамическом пределе большого ансамбля ``перемешивание'' большого набора случайных профилей в течение достаточно длительного времени приводит к равномерному распределению $W(\psi)=(2\pi)^{-1}$, что соответствует решению О-А. Таким образом, для реальных систем множество Отта-Антонсена оказывается не просто семейством частных решений, а аттрактором, и при описании динамики параметра порядка в системе можно полагаться на уравнение (\ref{eqap05}).

\newenvironment{ltrtr}{
\vspace{0.5\baselineskip}\noindent{\bf{Библиографический список}}\vspace{-0.5\baselineskip}
\begin{enumerate}
\partopsep=0pt\topsep=0pt\itemsep=1pt\parsep=0pt\parskip=0pt}
{\end{enumerate}}

\begin{ltrtr}
	
	\bibitem{Pikovsky-Rosenblum-Kurths-2001}
	{\em Пиковский А., Розенблюм М., Куртс Ю.}
	Синхронизация. Фундаментальное нелинейное явление.
	М: Техносфера, 2003. 496~с.
	Оригинал:
	{\em Pikovsky A., Rosenblum M., Kurths J.}
	Synchronization. A Universal Concept in Nonlinear Sciences.
	Cambridge: Cambridge University Press, 2003. 432~p.
	
	\bibitem{Crawford-1994}
	{\em Crawford J.D.}
	Amplitude expansions for instabilities in populations of globally-coupled oscillators //
	J.\ Stat.\ Phys. 1994. Vol.~74. P.~1047--1984.
	
	\bibitem{Strogatz-etal-2005}
	{\em Strogatz S.H., Abrams D.M., McRobie A., Eckhardt B., Ott E.},
	Theoretical mechanics: Crowd synchrony on the Millennium Bridge //
	Nature. 2005. Vol.~438. P.~43--44.
	
	\bibitem{Golomb-Hansel-Mato-2001}
	{\em Golomb D., Hansel D., Mato G.}
	Mechanisms of synchrony of neural activity in large networks //
	Handbook of Biological Physics, Volume 4: Neuroinformatics and Neural Modelling.
	Ed.\ by F.\ Moss and S.\ Gielen. Amsterdam: Elsevier, 2001. P.~887--968.
	
	\bibitem{Pikovsky-1984}
	{\em Пиковский А.С.}
	Синхронизация и стохастизация ансамбля автогенераторов внешним шумом //
	Изв.\ вузов. Радиофизика. 1984. Т.~27. С.~390--395.
	
	\bibitem{Mainen-Sejnowski-1995}
	{\em Mainen Z.F., Sejnowski T.J.}
	Reliability of spike timing in neocortical neurons //
	Science. 1995. Vol.~268. P.~1503--1506.
	
	\bibitem{Uchida-McAllister-Roy-2004}
	{\em Uchida A., McAllister R., Roy R.}
	Consistency of nonlinear system response to complex drive signals //
	Phys.\ Rev.\ Lett. 2004. Vol.~93. 244102.
	
	\bibitem{Grenfell-etal-1998}
	{\em Grenfell B.T., Wilson K., Finkenst\"adt B.F., Coulson T.N., Murray S., Albon S.D., Pemberton J.M., Clutton-Brock T.H., Crawley M.J.}
	Noise and determinism in synchronized sheep dynamics //
	Nature. 1998. Vol.~394. P.~674--677.
	
	\bibitem{Ritt-2003}
	{\em Ritt J.}
	Evaluation of entrainment of a nonlinear neural oscillator to white noise //
	Phys.\ Rev.\ E. 2003. Vol.~68, 041915.

    \bibitem{Goldobin-Pikovsky-2004}
    {\em Голдобин Д.С., Пиковский А.С.}
    О синхронизации периодических автоколебаний общим шумом //
    Изв.\ вузов. Радиофизика. 1984. Т.~47. С.~1013--1019.
	
	\bibitem{Teramae-Tanaka-2005}
	{\em Teramae J.N., Tanaka D.}
	Robustness of the Noise-Induced Phase Synchronization in a General Class of Limit Cycle Oscillators //
	Phys.\ Rev.\ Lett. 2004. Vol.~93. 204103.
	
	\bibitem{Goldobin-Pikovsky-2005a}
	{\em Goldobin D.S., Pikovsky A.S.}
	Synchronization of self-sustained oscillators by common white noise //
	Physica A. 2005. Vol.~351. №~1. P.~126--132.
	
	\bibitem{Goldobin-Pikovsky-2005b}
	{\em Goldobin D.S., Pikovsky A.}
	Synchronization and desinchronization of self-sustained oscillators by common noise //
	Phys.\ Rev.\ E. 2005. Vol.~71. 045201(R).
	
	\bibitem{Goldobin-Pikovsky-2006}
	{\em Goldobin D.S., Pikovsky A.}
	Antireliability of noise-driven neurons //
	Phys.\ Rev.\ E. 2006. Vol.~73. 061906.

	\bibitem{Malyaev-etal-2007}
	{\em Маляев В.С., Вадивасова Т.Е., Анищенко В.С.}
	Стохастический резонанс, стохастическая синхронизация и индуцированный шумом хаос в осцилляторе Дуффинга //
	Изв.\ вузов. Прикладная нелинейная динамика. 2007. Т.~15. №~5. С.~74--83.
	
	\bibitem{Wieczorek-2009}
	{\em Wieczorek S.}
	Stochastic bifurcation in noise-driven lasers and Hopf oscillators //
	Phys.\ Rev.\ E. 2009. Vol.~79. 036209.
	
	\bibitem{Goldobin-etal-2010}
	{\em Goldobin D.S., Teramae J.-N., Nakao H., Ermentrout G.-B.}
	Dynamics of Limit-Cycle Oscillators Subject to General Noise //
	Phys.\ Rev.\ Lett. 2010. Vol.~105. 154101.
	
	\bibitem{Goldobin-2014}
	{\em Goldobin D.S.}
	Uncertainty principle for control of ensembles of oscillators driven by common noise //
	Eur.\ Phys.\ J.\ ST. 2014. Vol.~223. №~4. P.~677--685.

	\bibitem{Goldobin-2014b}
	{\em Голдобин Д.С.}
	Принцип неопределенности для ансамблей осцилляторов с общим шумом //
	Вестник Пермского университета. Физика. 2014. Т.~27--28. Вып.~2--3. С.~33--41.
	
	\bibitem{Braun-etal-2012}
	{\em Braun W., Pikovsky A., Matias M.A., Colet P.}
	Global dynamics of oscillator populations under common noise //
	EPL. 2012. Vol.~99. 20006.
	
	\bibitem{Pimenova-etal-2016}
	{\em Pimenova A.V., Goldobin D.S., Rosenblum M., Pikovsky A.}
	Interplay of coupling and common noise at the transition to synchrony in oscillator populations //
	Sci.\ Rep. 2016. Vol.~6. 38518.
	
	\bibitem{Garcia-Alvarez-etal-2009}
	{\em Garc\'ia-\'Alvarez D., Bahraminasab A., Stefanovska A., McClintock P.V.E.}
	Competition between noise and coupling in the induction of synchronisation //
	EPL. 2009. Vol.~88. 30005.
	
	\bibitem{Nagai-Kori-2010}
	{\em Nagai K.H., Kori H.}
	Noise-induced synchronization of a large population of globally coupled nonidentical oscillators //
	Phys.\ Rev.\ E. 2010. Vol.~81. 065202.

    \bibitem{Wiener-1965}
    {\em Wiener N.}
    Cybernetics: Or Control and Communication in the Animal and the Machine. 2nd Ed.
    Cambridge (MA): MIT Press, 1965. 212~p.

    \bibitem{Martens-etal-2013}
    {\em Martens E.A., Thutupalli S., Fourri\`ere A., Hallatschek O.}
    Chimera states in mechanical oscillator networks //
    Proc.\ Natl.\ Acad.\ Sci. 2013. Vol.~110. P.~10563--10567.

    \bibitem{Temirbayev-etal-2012}
    {\em Temirbayev A.A., Zhanabaev Z.Z., Tarasov S.B., Ponomarenko V.I., Rosenblum M.}
    Experiments on oscillator ensembles with global nonlinear coupling //
    Phys.\ Rev.\ E. 2012. Vol.~85. 015204(R).

    \bibitem{Temirbayev-etal-2013}
    {\em Temirbayev A.A., Nalibayev Y.D., Zhanabaev Z.Z., Ponomarenko V.I., Rosenblum M.}
    Autonomous and forced dynamics of oscillator ensembles with global nonlinear coupling: An experimental study //
    Phys.\ Rev.\ E. 2013. Vol.~87. 062917.

	\bibitem{Watanabe-Strogatz-1994}
	{\em Watanabe S., Strogatz S.H.}
	Constant of Motion for Superconducting Josephson Arrays //
	Physica D. 1994. Vol.~74. P.~197--253.
	
	\bibitem{Pikovsky-Rosenblum-2008}
	{\em Pikovsky A., Rosenblum M.}
	Partially Integrable Dynamics of Hierarchical Populations of Coupled Oscillators //
	Phys.\ Rev.\ Lett. 2008. Vol.~101. 2264103.
	
	\bibitem{Marvel-Mirollo-Strogatz-2009}
	{\em Marvel S.A., Mirollo R.E., Strogatz S.H.}
	Identical phase oscillators with global sinusoidal coupling evolve by M\"obius group action //
	CHAOS. 2009. Vol.~19. 043104.
	
	\bibitem{Ott-Antonsen-2008}
	{\em Ott E., Antonsen T.M.}
	Low dimensional behavior of large systems of globally coupled oscillators //
	CHAOS. 2008. Vol.~18. 037113.
	
	\bibitem{Fletcher-Galerkin_method}
	{\em Флетчер К.}
	Численные методы на основе метода Галёркина. М: Мир, 1988. 352~с.

	\bibitem{Marvel-Strogatz-2009}
	{\em Marvel S.A., Strogatz S.H.}
	Invariant submanifold for series arrays of Josephson junctions //
	CHAOS. 2009. Vol.~19. 013132.

	\bibitem{Shinomoto-Kuramoto-1986}
    {\em Shinomoto Sh., Kuramoto Y.}
    Phase Transitions in Active Rotator Systems //
    Prog.\ Theor.\ Phys. 1986. Vol.~75. №~5. P.~1105--1110.
	
\end{ltrtr}


\begin{center}
    {\bf SYNCHRONIZATION IN KURAMOTO-SAKAGUCHI ENSEMBLES WITH COMPETING INFLUENCE OF COMMON NOISE AND GLOBAL COUPLING}

\vspace{5mm}
    {\it D.\ S.\ Goldobin$^{1,2}$, A.\ V.\ Dolmatova$^1$, M.\ Rosenblum$^{3,4}$, A.\ Pikovsky$^{3,4}$}

\vspace{5mm}
$^1$Institute of Continuous Media Mechanics UB RAS,
614013 Perm, Akad.\ Koroleva, 1\\
$^2$Perm State University,
614990 Perm, Bukireva, 15\\
$^3$University of Potsdam,
14476 Potsdam-Golm, Karl-Liebknecht-Str., 24/25\\
$^4$Nizhny Novgorod State University,
603950 Nizhny Novgorod, Gagarina, 23
\end{center}

\begin{abstract}
We study the effects of synchronization and desynchronization in ensembles of phase oscillators with the global Kuramoto-Sakaguchi coupling under common noise driving. Since the mechanisms of synchronization by coupling and by common noise are essentially different, their interplay is of interest. In the thermodynamic limit of large number of oscillators, employing the Ott-Antonsen approach, we derive stochastic equations for the order parameters and consider their dynamics for two cases: (i)~identical oscillators and (ii)~small natural frequency mismatch. For identical oscillators, the stability of the perfect synchrony state is studied; a strong enough common noise is revealed to prevail over a moderate negative (repelling) coupling and to synchronize the ensemble. An inequality between the states of maximal asynchrony (zero-value of the order parameter) and perfect synchrony; the former can be only weakly stable, while the latter can become adsorbing (the transition to the synchrony becomes unidirectional). The dependence of the temporal dynamics of the transition on the system parameters is investigated. For nonidentical oscillators the perfect synchrony state becomes impossible and an absorbing state disappears; on its place, only a weakly stable state of imperfect synchrony remains. A nontrivial effect of the divergence of individual frequencies of oscillators with different natural frequencies is revealed and studied for moderate repelling coupling; meanwhile, the order parameter remains non-small for this case. In Appendix we provide an introduction to the theories of Ott-Antonsen and Watanabe-Strogatz.

\noindent{\it Keywords:} Synchronization, stochastic processes, Kuramoto-Sakaguchi ensemble, Ott-Antonsen ansatz.

\end{abstract}

\renewenvironment{ltrtr}{
\vspace{0.5\baselineskip}\noindent{\bf{References}}\vspace{-0.5\baselineskip}
\begin{enumerate}
\partopsep=0pt\topsep=0pt\itemsep=1pt\parsep=0pt\parskip=0pt}
{\end{enumerate}}

\begin{ltrtr}
	
	\bibitem{_Pikovsky-Rosenblum-Kurths-2001}
	{\em Pikovsky A., Rosenblum M., Kurths J.}
	Synchronization. A Universal Concept in Nonlinear Sciences.
	Cambridge: Cambridge University Press, 2003. 432~p.
	
	\bibitem{_Crawford-1994}
	{\em Crawford J.D.}
	Amplitude expansions for instabilities in populations of globally-coupled oscillators //
	J.\ Stat.\ Phys. 1994. Vol.~74. P.~1047--1984.
	
	\bibitem{_Strogatz-etal-2005}
	{\em Strogatz S.H., Abrams D.M., McRobie A., Eckhardt B., Ott E.},
	Theoretical mechanics: Crowd synchrony on the Millennium Bridge //
	Nature. 2005. Vol.~438. P.~43--44.
	
	\bibitem{_Golomb-Hansel-Mato-2001}
	{\em Golomb D., Hansel D., Mato G.}
	Mechanisms of synchrony of neural activity in large networks //
	Handbook of Biological Physics, Volume 4: Neuroinformatics and Neural Modelling.
	Ed.\ by F.\ Moss and S.\ Gielen. Amsterdam: Elsevier, 2001. P.~887--968.
	
	\bibitem{_Pikovsky-1984}
    {\em Pikovsky A.S.}
	Synchronization and stochastization of the ensemble of autogenerators by external noise //
	Radiophys.\ Quantum Electron. 1984. Vol.~27. P.~576--581.
	
	\bibitem{_Mainen-Sejnowski-1995}
	{\em Mainen Z.F., Sejnowski T.J.}
	Reliability of spike timing in neocortical neurons //
	Science. 1995. Vol.~268. P.~1503--1506.
	
	\bibitem{_Uchida-McAllister-Roy-2004}
	{\em Uchida A., McAllister R., Roy R.}
	Consistency of nonlinear system response to complex drive signals //
	Phys.\ Rev.\ Lett. 2004. Vol.~93. 244102.
	
	\bibitem{_Grenfell-etal-1998}
	{\em Grenfell B.T., Wilson K., Finkenst\"adt B.F., Coulson T.N., Murray S., Albon S.D., Pemberton J.M., Clutton-Brock T.H., Crawley M.J.}
	Noise and determinism in synchronized sheep dynamics //
	Nature. 1998. Vol.~394. P.~674--677.
	
	\bibitem{_Ritt-2003}
	{\em Ritt J.}
	Evaluation of entrainment of a nonlinear neural oscillator to white noise //
	Phys.\ Rev.\ E. 2003. Vol.~68, 041915.

    \bibitem{_Goldobin-Pikovsky-2004}
    {\em Goldobin D.S., Pikovsky A.S.}
    Synchronization of periodic self-oscillations by common noise //
    Radiophys.\ Quantum Electron. 2004. Vol.~47. N.~10--11. P.~910--915.
	
	\bibitem{_Teramae-Tanaka-2005}
	{\em Teramae J.N., Tanaka D.}
	Robustness of the Noise-Induced Phase Synchronization in a General Class of Limit Cycle Oscillators //
	Phys.\ Rev.\ Lett. 2004. Vol.~93. 204103.
	
	\bibitem{_Goldobin-Pikovsky-2005a}
	{\em Goldobin D.S., Pikovsky A.S.}
	Synchronization of self-sustained oscillators by common white noise //
	Physica A. 2005. Vol.~351. N.~1. P.~126--132.
	
	\bibitem{_Goldobin-Pikovsky-2005b}
	{\em Goldobin D.S., Pikovsky A.}
	Synchronization and desinchronization of self-sustained oscillators by common noise //
	Phys.\ Rev.\ E. 2005. Vol.~71. 045201(R).
	
	\bibitem{_Goldobin-Pikovsky-2006}
	{\em Goldobin D.S., Pikovsky A.}
	Antireliability of noise-driven neurons //
	Phys.\ Rev.\ E. 2006. Vol.~73. 061906.

	\bibitem{_Malyaev-etal-2007}
	{\em Malyaev V.S., Vadivasova T.E., Anishchenko V.S.}
	Izvestija VUZ. Applied Nonlinear Dynamics. 2007. Vol.~15. N.~5. P.~74--83. (in Russian).
	
	\bibitem{_Wieczorek-2009}
	{\em Wieczorek S.}
	Stochastic bifurcation in noise-driven lasers and Hopf oscillators //
	Phys.\ Rev.\ E. 2009. Vol.~79. 036209.
	
	\bibitem{_Goldobin-etal-2010}
	{\em Goldobin D.S., Teramae J.-N., Nakao H., Ermentrout G.-B.}
	Dynamics of Limit-Cycle Oscillators Subject to General Noise //
	Phys.\ Rev.\ Lett. 2010. Vol.~105. 154101.
	
	\bibitem{_Goldobin-2014}
	{\em Goldobin D.S.}
	Uncertainty principle for control of ensembles of oscillators driven by common noise //
	Eur.\ Phys.\ J.\ ST. 2014. Vol.~223. N.~4. P.~677--685.

	\bibitem{_Goldobin-2014b}
	{\em Goldobin D.S.}
	An uncertainty principle for ensembles of oscillators driven by common noise //
	Bulletin of Perm University. Series: Physics. 2014. Vol.~27--28. N.~2--3. P.~33--41. (in Russian).
	
	\bibitem{_Braun-etal-2012}
	{\em Braun W., Pikovsky A., Matias M.A., Colet P.}
	Global dynamics of oscillator populations under common noise //
	EPL. 2012. Vol.~99. 20006.
	
	\bibitem{_Pimenova-etal-2016}
	{\em Pimenova A.V., Goldobin D.S., Rosenblum M., Pikovsky A.}
	Interplay of coupling and common noise at the transition to synchrony in oscillator populations //
	Sci.\ Rep. 2016. Vol.~6. 38518.
	
	\bibitem{_Garcia-Alvarez-etal-2009}
	{\em Garc\'ia-\'Alvarez D., Bahraminasab A., Stefanovska A., McClintock P.V.E.}
	Competition between noise and coupling in the induction of synchronisation //
	EPL. 2009. Vol.~88. 30005.
	
	\bibitem{_Nagai-Kori-2010}
	{\em Nagai K.H., Kori H.}
	Noise-induced synchronization of a large population of globally coupled nonidentical oscillators //
	Phys.\ Rev.\ E. 2010. Vol.~81. 065202.
	
    \bibitem{_Wiener-1965}
    {\em Wiener N.}
    Cybernetics: Or Control and Communication in the Animal and the Machine. 2nd Ed.
    Cambridge (MA): MIT Press, 1965. 212~p.

    \bibitem{_Martens-etal-2013}
    {\em Martens E.A., Thutupalli S., Fourri\`ere A., Hallatschek O.}
    Chimera states in mechanical oscillator networks //
    Proc.\ Natl.\ Acad.\ Sci. 2013. Vol.~110. P.~10563--10567.

    \bibitem{_Temirbayev-etal-2012}
    {\em Temirbayev A.A., Zhanabaev Z.Z., Tarasov S.B., Ponomarenko V.I., Rosenblum M.}
    Experiments on oscillator ensembles with global nonlinear coupling //
    Phys.\ Rev.\ E. 2012. Vol.~85. 015204(R).

    \bibitem{_Temirbayev-etal-2013}
    {\em Temirbayev A.A., Nalibayev Y.D., Zhanabaev Z.Z., Ponomarenko V.I., Rosenblum M.}
    Autonomous and forced dynamics of oscillator ensembles with global nonlinear coupling: An experimental study //
    Phys.\ Rev.\ E. 2013. Vol.~87. 062917.

	\bibitem{_Watanabe-Strogatz-1994}
	{\em Watanabe S., Strogatz S.H.}
	Constant of Motion for Superconducting Josephson Arrays //
	Physica D. 1994. Vol.~74. P.~197--253.
	
	\bibitem{_Pikovsky-Rosenblum-2008}
	{\em Pikovsky A., Rosenblum M.}
	Partially Integrable Dynamics of Hierarchical Populations of Coupled Oscillators //
	Phys.\ Rev.\ Lett. 2008. Vol.~101. 2264103.
	
	\bibitem{_Marvel-Mirollo-Strogatz-2009}
	{\em Marvel S.A., Mirollo R.E., Strogatz S.H.}
	Identical phase oscillators with global sinusoidal coupling evolve by M\"obius group action //
	CHAOS. 2009. Vol.~19. 043104.
	
	\bibitem{_Ott-Antonsen-2008}
	{\em Ott E., Antonsen T.M.}
	Low dimensional behavior of large systems of globally coupled oscillators //
	CHAOS. 2008. Vol.~18. 037113.
	
	\bibitem{_Fletcher-Galerkin_method}
	{\em Fletcher C.A.J.}
	Computational Galerkin methods. N.Y./Berlin/Heidelberg: Springer Verlag, 1984. 309~p.

	\bibitem{_Marvel-Strogatz-2009}
	{\em Marvel S.A., Strogatz S.H.}
	Invariant submanifold for series arrays of Josephson junctions //
	CHAOS. 2009. Vol.~19. 013132.

	\bibitem{_Shinomoto-Kuramoto-1986}
    {\em Shinomoto Sh., Kuramoto Y.}
    Phase Transitions in Active Rotator Systems //
    Prog.\ Theor.\ Phys. 1986. Vol.~75. No.~5. P.~1105--1110.
	
\end{ltrtr}

\begin{figure}[!ht]
\begin{tabular}{p{0.20\textwidth}p{0.76\textwidth}}
\noindent
  \includegraphics[width=0.20\textwidth]%
 {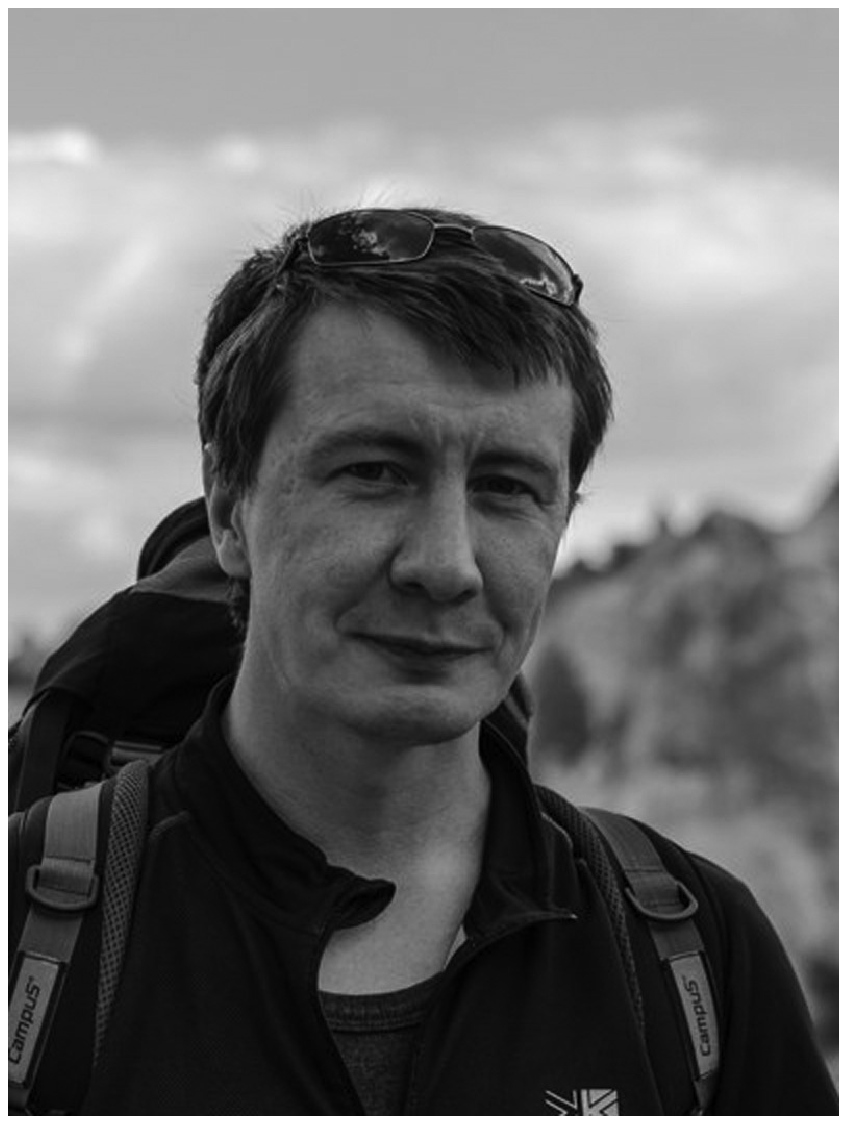}
 &
 \vspace{-42.5mm}
{\em Голдобин Денис Сергеевич} --- родился в Перми (1981),
окончил физический факультет Пермского государственного университета (2004).
В 2007 году защитил диссертации на соискание ученой степеней кандидата физико-математических наук по специальности <<Механика жидкости, газа и плазмы>> (ПермГУ)
и Dr.\ rer.\ nat.\ по теоретической физике (Университет Потсдама, Германия).
Работал в университетах Потсдама, Лестера и Перми и Институте механики сплошных сред УрО РАН (Пермь).
Область научных интересов --- вибрационные эффекты в гидродинамике, тепловая конвекция, статистическая физика, нелинейная динамика, моделирование геофизических процессов.
Автор более 50 научных статей по указанным выше направлениям.

E-mail: Denis.Goldobin@gmail.com

Группа динамики геологических систем, ИМСС УрО РАН

614013 Пермь
\end{tabular}
\end{figure}

\begin{figure}[!ht]
\begin{tabular}{p{0.20\textwidth}p{0.76\textwidth}}
\noindent
  \includegraphics[width=0.20\textwidth]%
 {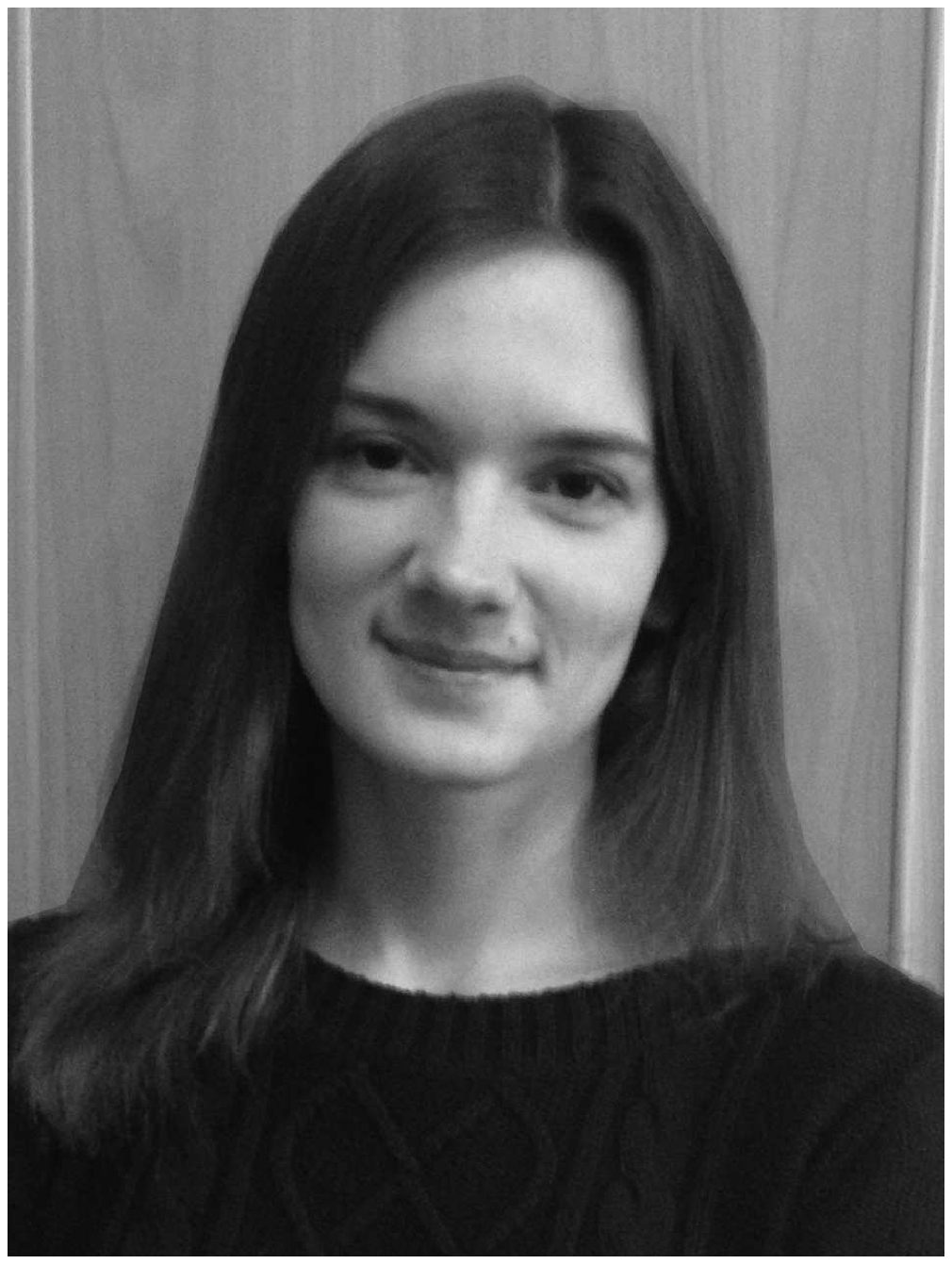}
 &
 \vspace{-42.5mm}
{\em Долматова (Пименова) Анастасия Владимировна} --- родилась в Перми (1990),
окончила физический факультет Пермского государственного университета (2012).
Защитила диссертацию на соискание ученой степеней кандидата физико-математических наук по специальности <<Механика жидкости, газа и плазмы>> (ИМСС УрО РАН, Пермь, 2016).
Работает в Институте механики сплошных сред УрО РАН (Пермь).
Область научных интересов --- механика жидкости и газа, статистическая физика, моделирование геофизических процессов.
Автор 17 научных статей по указанным выше направлениям.

E-mail: Anastasiya.Pimenova@gmail.com

Группа динамики геологических систем, ИМСС УрО РАН

614013 Пермь
\end{tabular}
\end{figure}

\begin{figure}[!ht]
\begin{tabular}{p{0.20\textwidth}p{0.76\textwidth}}
\noindent
  \includegraphics[width=0.20\textwidth]%
 {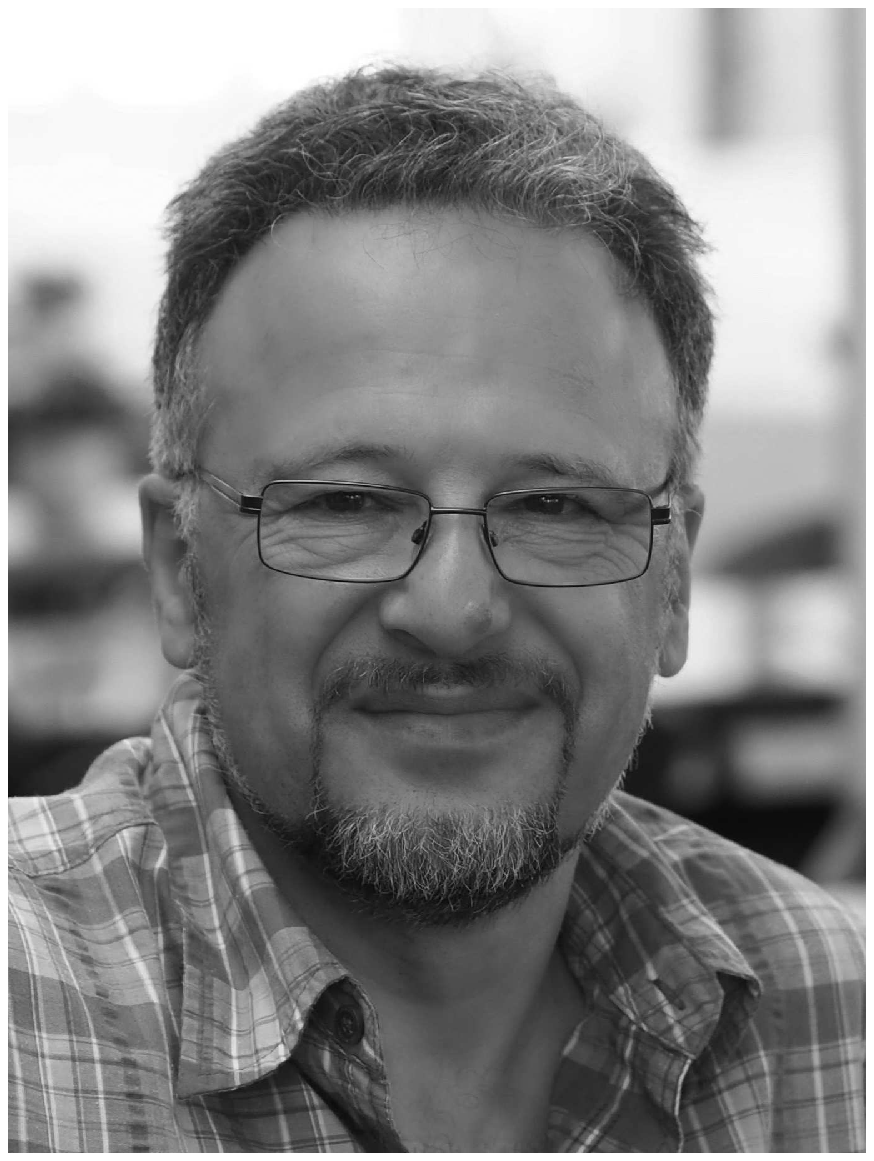}
 &
 \vspace{-42.5mm}
{\em Розенблюм Михаил Григорьевич} --- родился в Москве (1958), окончил Московский государственный педагогический институт (1980).
После окончания работал в Институте машиноведения АН, затем в университетах Бостона (BU) и Потсдама.
Защитил диссертацию на соисканиеученой степени кандидата физико-математических наук (СГУ, 1990).
Область научных интересов --- нелинейная динамика, синхронизация, анализ данных.

E-mail: mros@uni-potsdam.de

Department of Physics and Astronomy, Potsdam University

14476 Potsdam-Golm, Germany
\end{tabular}
\end{figure}

\begin{figure}[!ht]
\begin{tabular}{p{0.20\textwidth}p{0.76\textwidth}}
\noindent
  \includegraphics[width=0.20\textwidth]%
 {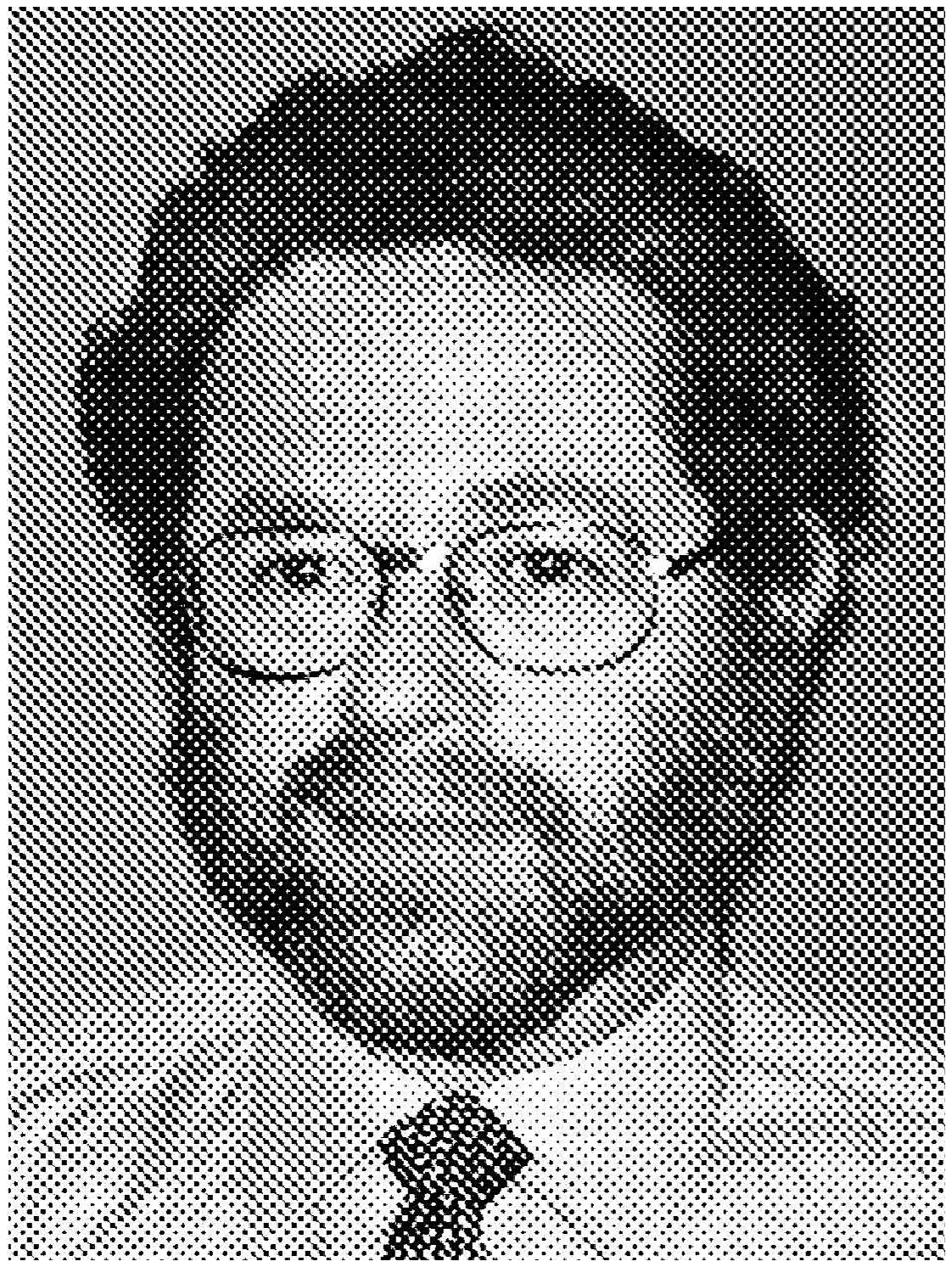}
 &
 \vspace{-42.5mm}
{\em Пиковский Аркадий Самуилович} --- родился в Горьком (1956), окончил Горьковский государственный университет (1977). После окончания работал в Институте прикладной физики АН, затем в университетах Вупперталя и Потсдама. Защитил диссертацию на соискание ученой степени кандидата физико-математических наук (ГГУ, 1982). Область научных интересов --- нелинейная динамика, статистическая физика и теория хаоса.

E-mail: pikovsky@uni-potsdam.de

Department of Physics and Astronomy, Potsdam University

14476 Potsdam-Golm, Germany
\end{tabular}
\end{figure}

\clearpage

\begin{figure}[!t]
Fig.~1:
Dependence of $\la\sin^2\varPhi\ra$ on $\varOmega_\mu/\sigma^2$ determines the Lyapunov exponent $\lambda$ (see equarion~(11)). The black solid line represents exact solution~(13), the dashed line shows the Galerkin's approximation (17)--(18), the dash-dotted line shows asymptotic expansion~(14). For the exact solution and Galerkin's approximation, $\la\sin^2\varPhi\ra$ tends to a nonzero finite value (approximately $0.2$) as $\varOmega_\mu/\sigma^2\to0$.
\end{figure}

\begin{figure}[!t]
Fig.~2:
The dependence of the order parameter $\la{J}\ra$ on $\mu\cos\beta/\sigma^2$~(see equation~(43)) is plotted with solid lines for $\gamma=1$, $10^{-1}$, $10^{-2}$, $10^{-3}$, $10^{-4}$ (from bottom to top), the dashed line tending to infinity at $\mu\cos\beta/\sigma^2=-1$, corresponds to $\gamma=0$.
\end{figure}

\begin{figure}[!t]
Fig.~3:
Dependence of the average frequency $\la\dot\theta\ra$ on the natural frequency mismatch $\omega$ for $\beta=0$ (a), $\pi/4$ (b), $\pi/2$ (c), and $3\pi/4$ (d). The results of numerical calculation by means of continuous fractions at $\gamma=0.01$ are plotted in (a), (b), (d) with circles, diamonds, up- and down-pointed triangles, squares ($\mu/\sigma^2=0.4$, $0.2$, $0$, $-0.2$, $-0.4$, respectively). The results in (c) are for $\mu/\sigma^2=0.8$, $0.4$, $0$, $-0.4$, $-0.8$. The black solid lines show the results for $\gamma=0$ and specified values of $\mu/\sigma^2$; the dashed lines indicate the slope $1+2\mu\cos\beta/\sigma^2$, corresponding to asymptotic behaviour $\la\dot\theta\ra\sim\omega^{1+2\mu\cos\beta/\sigma^2}$. In graphs (b)--(d) some curves cross zero not at $\omega=0$ --- in the log--log scale one observes this as horizontal plateaux (if $\la\dot\theta\ra_{\omega=0}>0$) or vertical drops (if $\la\dot\theta\ra_{\omega=0}<0$); in Fig.~4 the dependencies are plotted with bias so, that they cross zero at the origin.
\end{figure}

\begin{figure}[!t]
Fig.~4:
The dependence of the biased average frequency $\la\dot\theta\ra-\la\dot\theta\ra_{\omega=0}$ on the natural frequency mismatch $\omega$ at $\gamma=0.01$ and $\beta=\pi/4$ (a), $\pi/2$ (b), $3\pi/4$ (c); notations are the same as in Fig.~3
\end{figure}

\begin{figure}[!t]
Fig.~5:
The results of numerical simulation for the ensemble of $41$ oscillators with Gaussian distribution of natural frequencies. The dynamics of the phase difference of two oscillators in the ensemble for different levels of the coupling strength.
\end{figure}

\end{document}